\documentclass[%
 aip,
 amsmath,amssymb,
 reprint,%
]{revtex4-1}

\usepackage{graphicx}
\usepackage{dcolumn}
\usepackage{bm}

\usepackage[utf8]{inputenc}
\usepackage[T1]{fontenc}
\usepackage{mathptmx}
\usepackage{etoolbox}
\usepackage{braket}
\usepackage{physics}
\usepackage[normalem]{ulem}
\usepackage{tikz}
\usepackage{siunitx}
\usepackage{soul}
\usepackage{comment}

\usepackage{hyperref}
\hypersetup{colorlinks=true, linkcolor=black, citecolor=blue, urlcolor=blue}

\makeatletter
\def\@email#1#2{%
 \endgroup
 \patchcmd{\titleblock@produce}
  {\frontmatter@RRAPformat}
  {\frontmatter@RRAPformat{\produce@RRAP{*#1\href{mailto:#2}{#2}}}\frontmatter@RRAPformat}
  {}{}
}%

\makeatother
\begin{document}


\title{$^{88}$Sr$^+$ Ion Trap Apparatus for Generating 408\,nm Photons}
\author{Jianlong Lin}
\affiliation{Department of Electrical and Computer Engineering, University of Illinois Urbana-Champaign, Urbana, IL, USA}
\author{Mari Cieszynski}%
\author{William Christopherson}
\author{Darman Khan}
\author{Lintao Li}
\author{Elizabeth Goldschmidt}
\author{Brian DeMarco}
\email{bdemarco@illinois.edu}
\affiliation{Department of Physics, University of Illinois Urbana-Champaign, Urbana, IL, USA
}%

\date{\today}

\begin{abstract}

We describe a $^{88}$Sr$^+$ ion trap apparatus with the capability to produce high-quality 408\,nm photons aimed at distributed quantum computing and networking applications. This instrument confines ion chains using a surface electrode trap with a two-dimensional magneto-optical trap as an atomic source. Several laser systems spanning 400--1100\,nm are used to achieve high fidelity state preparation and readout. Photons are produced via the decay of an exited state, which is accessed using a custom 408\,nm laser system that produces 150\,ps optical pulses using non-linear photonics. We demonstrate single photon production through a Hanbury Brown-Twiss measurement for one to six ions.
\end{abstract}

\maketitle

\section{\label{sec:level1}Introduction}

Establishing high-fidelity entanglement between separated quantum bits (qubits) in a quantum network generally requires reliable, coherent control of quantum light--matter interfaces. Quantum information must be converted from a ``stationary'' qubit, held in one or more matter-based qubits, into a ``flying'' qubit encoded in a single photon or other quantum state of light~\cite{DiVincenzo_computing_2000, Kimble_internet_2008}. One promising paradigm for such an interface is a single emitter with internal spin structure that can be induced to emit a single photon whose state is entangled with its spin state. The potential functionality of such a system expands greatly when the emitter can be part of a larger quantum register comprised of multiple qubits, such as neutral atoms or trapped atomic ions~\cite{Covey_network_2023, Main_network_2025}.

Another important feature required for high-fidelity entanglement distribution is that the emitter produces photons in a well-defined quantum state, which can be difficult to achieve in practice. Many solid-state emitters, for example, produce light with a multiphoton background~\cite{flagg2012dynamics}, with excess spectral broadening due to dephasing during the emission time~\cite{Esmann_solid_2024, Liu_nv_2021}, or such that subsequent photons are not spectrally identical~\cite{Andrini_photons_2024, Waltrich_interference_2023}.

Here, we demonstrate the development of an instrument to trap $^{88}$Sr$^+$ ions as a high quality light--matter interface to 408\,nm photons. We show suppression of multiphoton emission to the level of $10^{-3}$.  While we do not directly measure the spectral purity in this work, we expect to be able to achieve high indistinguishability due to the nature of trapped ions being identical in vacuum~\cite{Bruzewicz_computing_2019}. The architecture is compatible with implementing quantum circuits for quantum information processing after entanglement distribution via single and two-qubit gates within a register~\cite{Main_network_2025}.


\section{Ion Trapping Apparatus}

In this section, we describe the design and implementation of the instrument that is used to trap $^{88}$Sr$^{+}$ atomic ions, prepare their quantum state, and produce fast 408\,nm excitation pulses to generate photons that are collected via an imaging system. The equipment includes an ultra-high vacuum system, continuous-wave lasers and associated optical systems, imaging optics and detectors, and a custom 150\,ps pulsed laser system.



\subsection{\label{sec:level2}Vacuum system}

The ultra-high vacuum (UHV) system we have built is designed to provide an environment for chip-based ion trap experiments. We use a surface electrode trap designed and fabricated by the Massachusetts Institute of Technology Lincoln Laboratory (MIT~LL) Quantum Computing Laboratory team. This trap can confine ions in three separate regions situated along the axial direction and is similar to the device described in Ref.~\citenum{sage_loading_2012}. Our variation is composed of niobium conducting electrodes deposited on a sapphire substrate, with 11-electrode pairs along the axial direction of the trap. We trap ions in just one zone for the experiments described in this manuscript. Using a simulation of the trap potential, we selected a set of RF and DC voltages to produce approximately 1.3\,MHz axial and 4\,MHz radial trap frequencies, with the radial principal axes tilted by approximately 30 degrees relative to the plane of the trap.

\begin{figure}
    \centering
    \includegraphics[width=1.0\columnwidth]{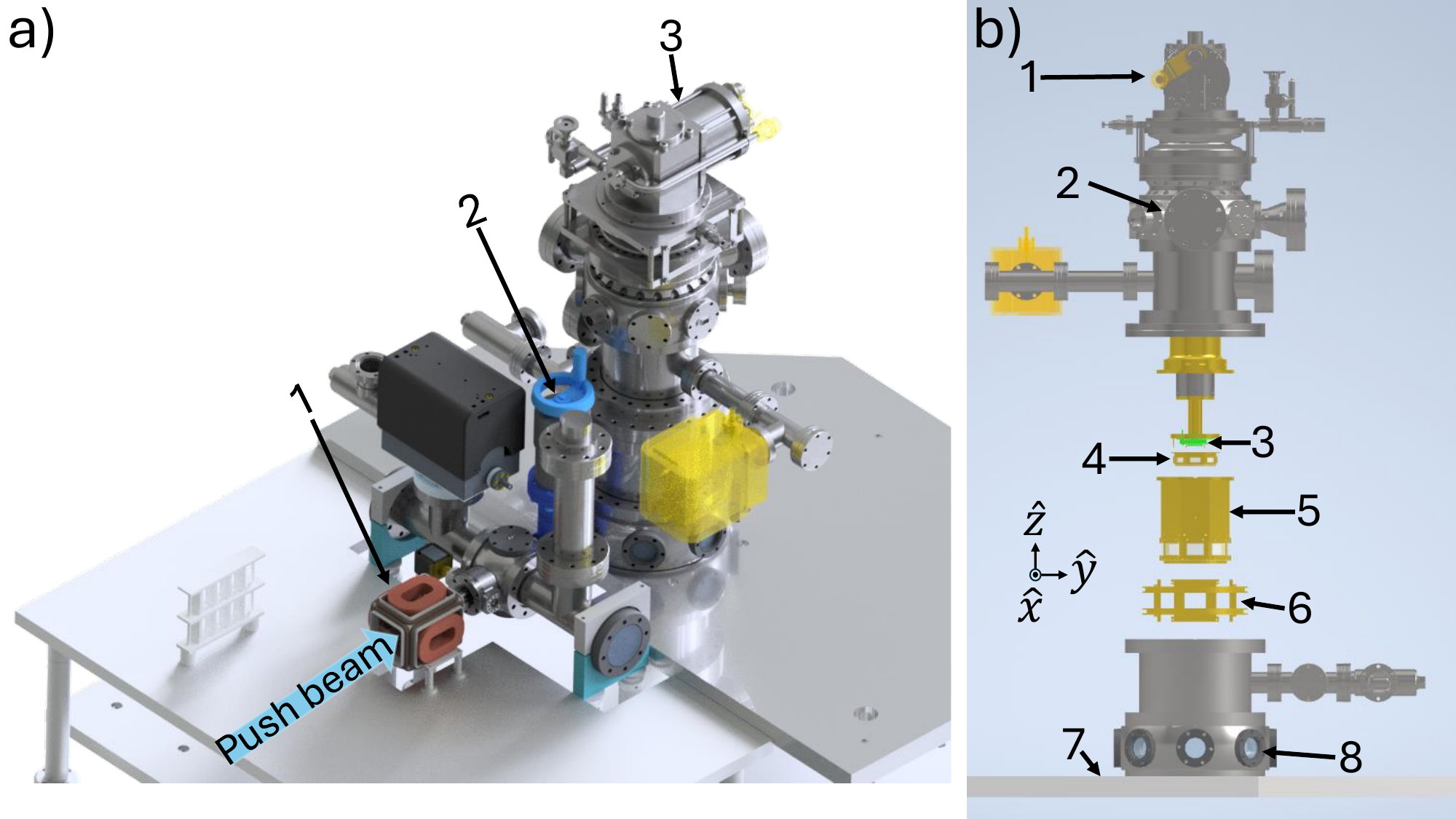}
    \caption{\label{fig:vacuum_system_and_cryostat} Vacuum system design. (a) A rendered model showing the full assembly: (1) 2D MOT glass cell and coils, (2) gate valve, and (3) cold head. (b) An exploded view showing the inside of the vacuum system surrounding the cryostat: (1) cold head, (2) feed-throughs for wiring, (3) trap assembly, (4) 4\,K radiation shield, (5) 50\,K radiation shield, (6) magnetic field coils, (7) breadboard with 6.2~inch bore for imaging, and (8) viewports for beam delivery. The wiring has been excluded for simplicity.}
\end{figure}

The UHV system, shown in Fig.~\ref{fig:vacuum_system_and_cryostat}, is composed of two chambers separated by a gate valve. In one chamber we generate an isotope-selective, two-dimensional magneto-optical trap (2D MOT) of $^{88}$Sr atoms, and in the other we trap and manipulate $^{88}$Sr$^+$ atomic ions. The 2D MOT is used to enable fast loading from a pre-cooled source of atoms~\cite{sage_loading_2012,bruzewicz_scalable_2016}. This approach avoids depositing unwanted material onto the trap surface. The 2D MOT uses a heated metal source to create a vapor of Sr atoms and is a $^{88}$Sr variation of the design described in Ref.~\onlinecite{huie_repetitive_2023}. After generating the 2D MOT, a laser beam pushes the atoms to the second chamber, where they are ionized and trapped. 

The UHV environment is maintained using a Gifford-McMahon cryocooler (Lakeshore Cryotronics low-vibration SHI-4XG-UHV-15). The cryocooler enables a wide range of materials to be used in the vacuum system and rapid swapping of components without the need to bake at high temperature to achieve UHV. Also, the ion trap chip is connected to the cryocooler cold finger, which reduces the anomalous ion motion heating rate~\cite{PhysRevA.89.012318}. A temperature of approximately 4\,K is achieved at our trap at the cryocooler base temperature. The design of this system is a variation of that described in Ref.~\citenum{sage_loading_2012}. Our system, as shown in Fig.~\ref{fig:vacuum_system_and_cryostat}, includes two changes compared with previous designs: an optical breadboard that supports the vacuum system, and vector magnetic field coils that are located inside the vacuum system.

By mounting the vacuum system to a custom breadboard, the ion addressing optics are mechanically coupled to the vacuum system, and thereby to the trap. This design minimizes the intensity and phase fluctuations that individual ions experience from the addressing beams due to mechanical vibrations. The optics for laser beam delivery are mounted to this breadboard and direct each beam through a viewport (Kurt J. Lesker VPZL-275) to manipulate ions located beneath the trap. This custom 1.5~inch thick aluminum optical breadboard also has a 6.2~inch bore cut out to provide optical access for the imaging system. 

Three-dimensional (3D) in-vacuo magnetic field coils provide a static field of approximately 0.5\,mT; similar designs have previously been used to provide a magnetic field along just one direction~\cite{sage_loading_2012, shi_long-lived_2024}. A 3D model of the coils, along with the principal directions of the magnetic field, are shown in Fig.~\ref{fig:vacuum_system_and_cryostat}. These coils provide 3D control over the magnetic field at the ions in a compact package. The magnetic field coils are composed of custom electro-polished, gold-plated, oxygen-free high thermal conductivity (OFHC) copper coil holders wound with polyimide-coated copper magnet wire. The coils are encapsulated in Stycast~2850FT with Catalyst~9. The coils that generate magnetic fields in the $\hat{x}$ and $\hat{y}$ directions (i.e., in the plane of the ion-trap chip) are wound with 108 turns of 22~AWG wire, and the pair that generates a magnetic field in $\hat{z}$ (i.e., orthogonal to the plane of the surface trap) is wound with 56 turns of 20~AWG wire. Producing a 0.5\,mT field requires 0.8\,A ($\hat{z}$ pair) and 1.7\,A ($\hat{x}$ and $\hat{y}$ pairs) currents. The coil holders connect directly to the 50\,K radiation shield, with enough surface area to ensure sufficient thermal contact. For our setup, the magnetic field is configured to be pointing in the $-\hat{y}$ direction indicated in Fig.~\ref{fig:vacuum_system_and_cryostat}(b) at the trapping location.

\subsection{\label{sec:BeamsAndImaging}Laser systems for trapping, state preparation, and measurement}

\begin{figure}
    \centering
    \includegraphics[width=1.0\columnwidth]{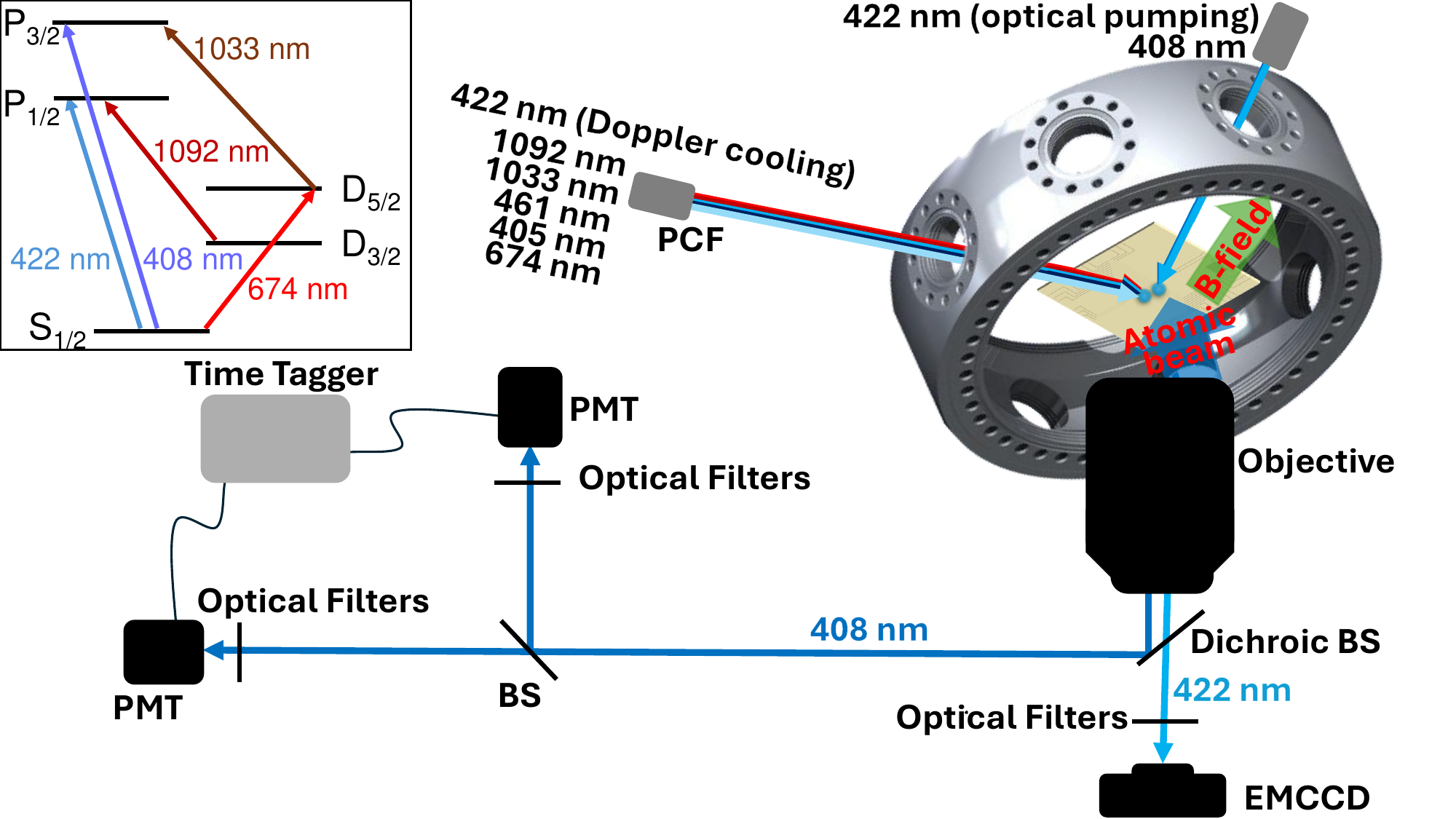}
    \caption{\label{fig:imaging_and_beams} Diagram showing laser-beam geometry used to manipulate trapped ions and the imaging system that collects fluorescence. Two optical systems deliver focused laser light to the trapping zone. An NA\,$\approx$\,0.48 objective collects ion fluorescence and directs it to a Hanbury-Brown and Twiss (HBT) measurement setup. Inset: energy level diagram of $^{88}$Sr$^+$ showing the laser wavelengths used to address transitions between quantum states.}
\end{figure}

Several laser beams at different wavelengths are needed to trap ions and for state preparation and measurement. A diagram of the laser beam geometry and energy levels of $^{88}$Sr$^+$ is shown in Fig.~\ref{fig:imaging_and_beams}. The acousto-optic modulators (AOMs) used to turn these laser beams on and off during data collection are not shown.

To trap ions, we follow the method described in Ref.~\onlinecite{sage_loading_2012}. A 461\,nm beam resonant with the $^{1}S_{0}\leftrightarrow{}^{1}P_{1}$ transition in neutral $^{88}$Sr pushes atoms from the 2D MOT to the ion trap chip. An additional 461\,nm beam (20\,{\textmu}W, linearly polarized relative to the magnetic field) combines with a 405\,nm beam (800\,{\textmu}W, $\sigma^{+/-}$-polarized) at the trapping region to photoionize $^{88}$Sr.  The push beam has a waist (i.e., $1/e^2$ radius) of 1\,mm, which is large enough to supply ions to any trapping region. To selectively load to the desired zone, the photoionization beam, which propagates at a 90/45 degree angle to the neutral atom beam/axial direction, is focused to a waist of approximately 30\,{\textmu}m. The 461\,nm beams used for the 2D-MOT, push beam, and photoionization are produced by an external cavity diode laser (ECDL, Toptica DL pro HP 461). The 405\,nm second-stage photoionization beam is produced by a Toptica iBEAM-SMART-405-S-HP.



The 0.5\,mT magnetic field lifts the degeneracy of the $S_{1/2}$ manifold to the $\ket{\downarrow}\equiv\ket{S_{1/2,}, ~m_j=-1/2}$ and $\ket{\uparrow}\equiv\ket{S_{1/2,}, ~m_j=+1/2}$ sublevels, which is referred to as the Zeeman qubit. The ions are prepared for single photon emission at 408\,nm on the closed $\ket{\downarrow}\leftrightarrow \ket{P_{3/2,}, ~m_j=-3/2}$ transition, first by Doppler cooling and optical pumping. Doppler cooling is implemented using a few~{\textmu}W of $\pi$-polarized 422\,nm light to drive the $S_{1/2}\leftrightarrow P_{1/2}$ transition. The ions are then optically pumped to the $\ket{\downarrow}$ state using a few~{\textmu}W of $\sigma^{-}$-polarized 422\,nm light. The 422\,nm light is derived from an external cavity diode laser (ECDL, Toptica DL pro HP 420). Light to quench atoms from the metastable $D_{3/2}$ and $D_{5/2}$ states, which have a small branching ratio for decay from the $P_{1/2}$ and $P_{3/2}$ states, is produced by a 1092\,nm laser (Toptica DL pro using diode LD-1120-0100-AR-2) and a 1033\,nm laser (Toptica DL pro using diode LD-1060-0200-AR-2). Approximately 20\,{\textmu}W of 1092\,nm $\sigma^{+/-}$-polarized light and 3\,{\textmu}W of 1033\,nm $\sigma^{+/-}$-polarized light is used for quenching.


To optimize and characterize the state preparation and measurement (SPAM) fidelity, we use spin-state selective readout to distinguish between $\ket{\downarrow}$ and $\ket{\uparrow}$ via fluorescence on the $S_{1/2}\leftrightarrow P_{1/2}$ transition~\cite{Keselman_state_detection_2011}. The population in $\ket{\downarrow}$ is first shelved to the metastable $D_{5/2}$ manifold using a narrow-linewidth (<10\,Hz) laser at 674\,nm so that the $\ket{\downarrow}$ state does not fluoresce under exposure to 422\,nm light. We use approximately 1\,mW of 674\,nm light polarized orthogonally to the magnetic field to drive $\Delta m_j=\pm2$ transitions. To estimate the fidelity of optical pumping and shelving separately, we apply a double-shelving technique. We first measure the combined optical pumping, shelving, and readout fidelities after applying separate experimental sequences to only shelve to $\ket{D_{5/2,}, ~m_j=-5/2}$ or $\ket{D_{5/2,}, ~m_j=+3/2}$ using $\pi$ pulses. We then employ a sequence that applies two $\pi$ pulses, consecutively resonant with the $\ket{\downarrow}\rightarrow\ket{D_{5/2,}, ~m_j=-5/2}$ and $\ket{\downarrow}\rightarrow\ket{D_{5/2,}, ~m_j=+3/2}$ transitions, and measure the combined fidelity of this double-shelving sequence. By factorizing the optical pumping and individual shelving probabilities, and estimating the overall fidelity for each sequence, we isolate our fidelities for optical pumping and for the $\ket{\downarrow}\rightarrow\ket{D_{5/2,}, ~m_j=-5/2}$ and $\ket{\downarrow}\rightarrow\ket{D_{5/2,}, ~m_j=+3/2}$ shelving transitions. Here, we neglect the small decay probability from $D_{5/2}$ to $S_{1/2}$. We estimate a combined optical pumping and readout fidelity of $>$\,99\,\%, optimized by adjusting the magnetic field coil currents to align the static magnetic field to be parallel with the optical pumping beam and by optimizing the circular polarization purity. Because the 422\,nm optical pumping beam and 408\,nm beam used for single-photon schemes co-propagate, the magnetic field alignment optimizes the $\sigma^{-}$ polarization of both beams.



The laser wavelengths are stabilized in several ways. The 422\,nm laser is locked to the $5P_{1/2}\leftrightarrow6P_{1/2}$ transition in $^{85}\mathrm{Rb}$ via saturated absorption spectroscopy~\cite{Shiner2006} in a heated rubidium vapor cell (Thorlabs GC25075-RB with GCH25R). An AOM (Isomet M1250-T250L-0.45) in a double-passed configuration is used to shift the 422\,nm beam to near resonance. The 461\,nm, 1033\,nm, 1092\,nm, and 816\,nm ECDLs are frequency-stabilized using measurement by a wavelength meter (HighFinesse WS8-10) and feedback to the ECDL gratings. The 674\,nm light is produced by a custom narrow-linewidth laser system from Stable Laser Systems (SLS-674-300-1) composed of an ECDL (Toptica DL pro 670) stabilized to a high-finesse optical cavity via the Pound-Drever-Hall technique. The light from a laser diode (Toptica Eagleyard EYP-RWE-0670-00703-1000-SOT02-0000) injection locked to the 300\,{\textmu}W transmitted output of the cavity is amplified using a tapered amplifier (TA) (BoostTA Pro-3V0). Two AOMs (Gooch \& Housego 3200-1214) are used to shift this beam to resonance.


The photoionization, Doppler cooling, shelving, and metastable state repumping beams are joined together using dichroic beamsplitters into a constant mode-field diameter fiber photonic-crystal fiber (PCF) (NKT Photonics aeroGUIDE-5-PM).  The output of this fiber is focused on the ions using silver-coated parabolic mirrors. This is an adaptation of a design used by MIT LL~\cite{beam_delivery_MIT_2024} and is constructed to focus the laser beams with different wavelengths to the same location. The waists of these beams at the focus ranges from 31\,{\textmu}m to 43\,{\textmu}m. To characterize the performance, we measure the location of the foci at 405\,nm and 1092\,nm ex-situ. Along the optical axis, the foci are shifted by 1.4\,mm, and perpendicular to the axis they are shifted by less than 9\,{\textmu}m. These distances are small compared with the shortest Rayleigh length and the beam waists, respectively. The $\sigma^{-}$-polarized 422\,nm optical pumping and 408\,nm excitation pulse beams (see Section~\ref{sec:PulsedSystem}) are delivered through a single-mode, polarization-maintaining fiber and similar focusing optics, with a polarizing beam splitter and a half- and a quarter- waveplate placed before the focusing mirror. 



\subsection{\label{sec:Imaging}Imaging system for photon collection and ion readout}

The setup for ion imaging and photon collection and state readout is shown in Fig.~\ref{fig:imaging_and_beams}. To collect fluorescence at 422\,nm for readout and single photons at 408\,nm from the ions, we use a 0.48 numerical aperture (NA) objective (Sill SS6ASS2258)~\cite{sage_loading_2012}, which was custom-designed for these wavelengths. After the objective, a dichroic beamsplitter (Semrock FF414-Di01) directs the 422\,nm and 408\,nm light along separate imaging paths. The 422\,nm light is steered to an Electron-Multiplying Charge-Coupled Device (EMCCD) camera (Andor iXon Ultra 888) for state readout between ground states of the $S_{1/2}$ manifold and to check if ions are trapped. The 408\,nm light emitted from the ion is sent through a Hanbury Brown and Twiss (HBT) setup to measure single-photon purity. The HBT setup employs photomultiplier tube (PMT) detectors (Hamamatsu H10682-210) along with a time-correlated single-photon counting module (Swabian Time Tagger Ultra) for coincidence measurements. Spectral filters ensure that unwanted light does not reach the EMCCD (one Semrock FF01-420/5-25 and one Semrock BSP01-633R-25) or PMTs (two Semrock ET402/15x and one Semrock FF01-492/SP-25).


\subsection{\label{sec:PulsedSystem}408\,nm pulsed laser system}

To generate single photons for use as a spin-photon interface and for future entanglement schemes, we excite the ion from $\ket{\downarrow}$ to the $\ket{e}\equiv\ket{P_{3/2,},~m_j=-3/2}$ state using a custom 408\,nm pulsed-laser system. To ensure a high probability of producing only a single photon, the excitation pulse must be  shorter than the $\tau = 6.99$\,ns excited state lifetime~\cite{Sansonetti2012, Brownnutt2007}. We have constructed a laser system capable of producing 150\,ps 408\,nm pulses to satisfy this criterion.

\begin{figure}
    \centering
    \includegraphics[width=1.0\columnwidth]{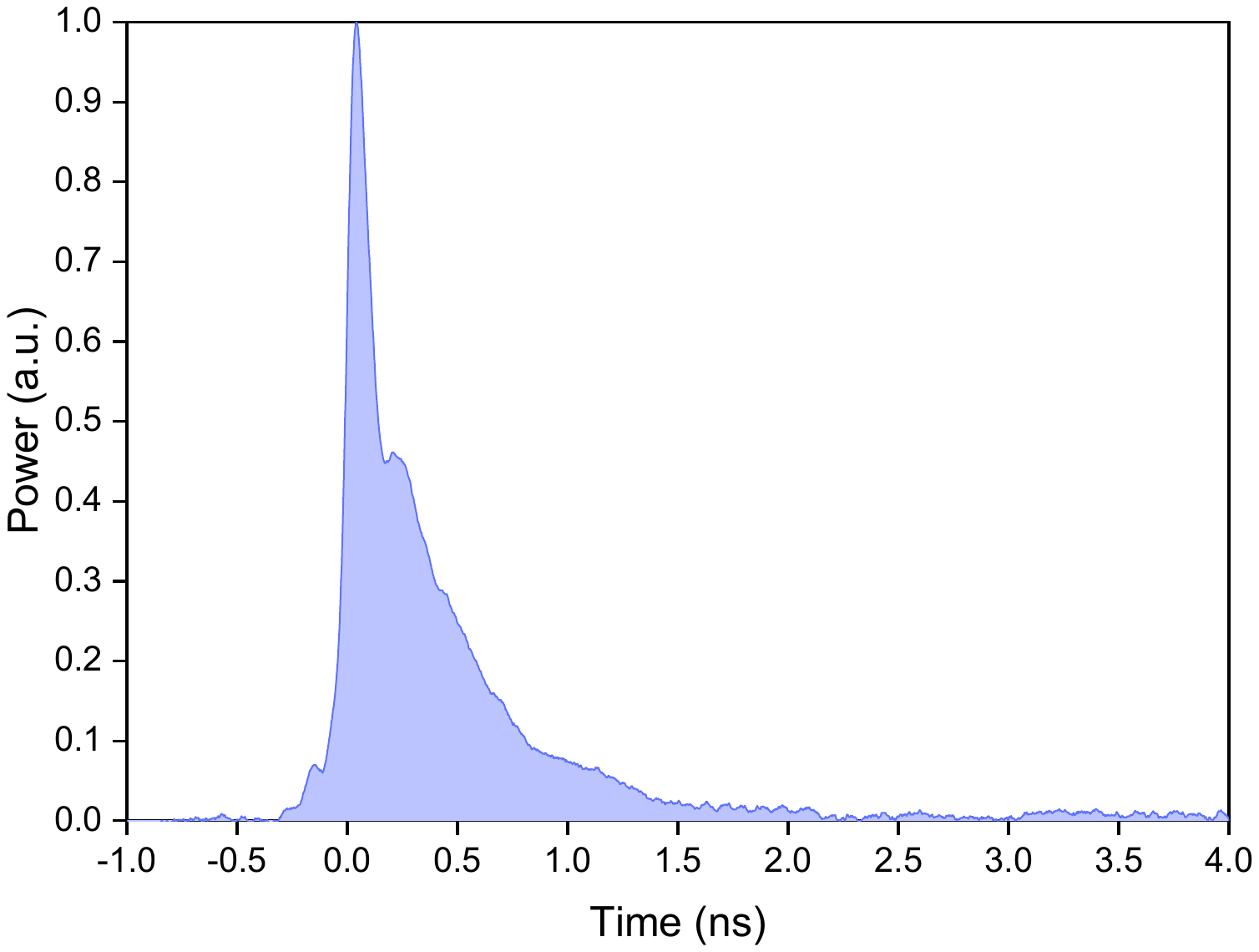}
    \caption{\label{fig:pulsed_system_output} Normalized output of the 408\,nm pulsed system, measured using an oscilloscope and photodiode with 12.5\,GHz bandwidths. The pulse has a FWHM of 148\,$\pm$\,3\,ps, set by the electronic pulse generation circuit. The RF amplifier (Analog Devices EVAL-HMC998APM5E) used to drive the EOM broadens the tail at the end of the pulse.}
\end{figure}

\begin{figure*}
    \centering
    \includegraphics[width=1.0\textwidth]{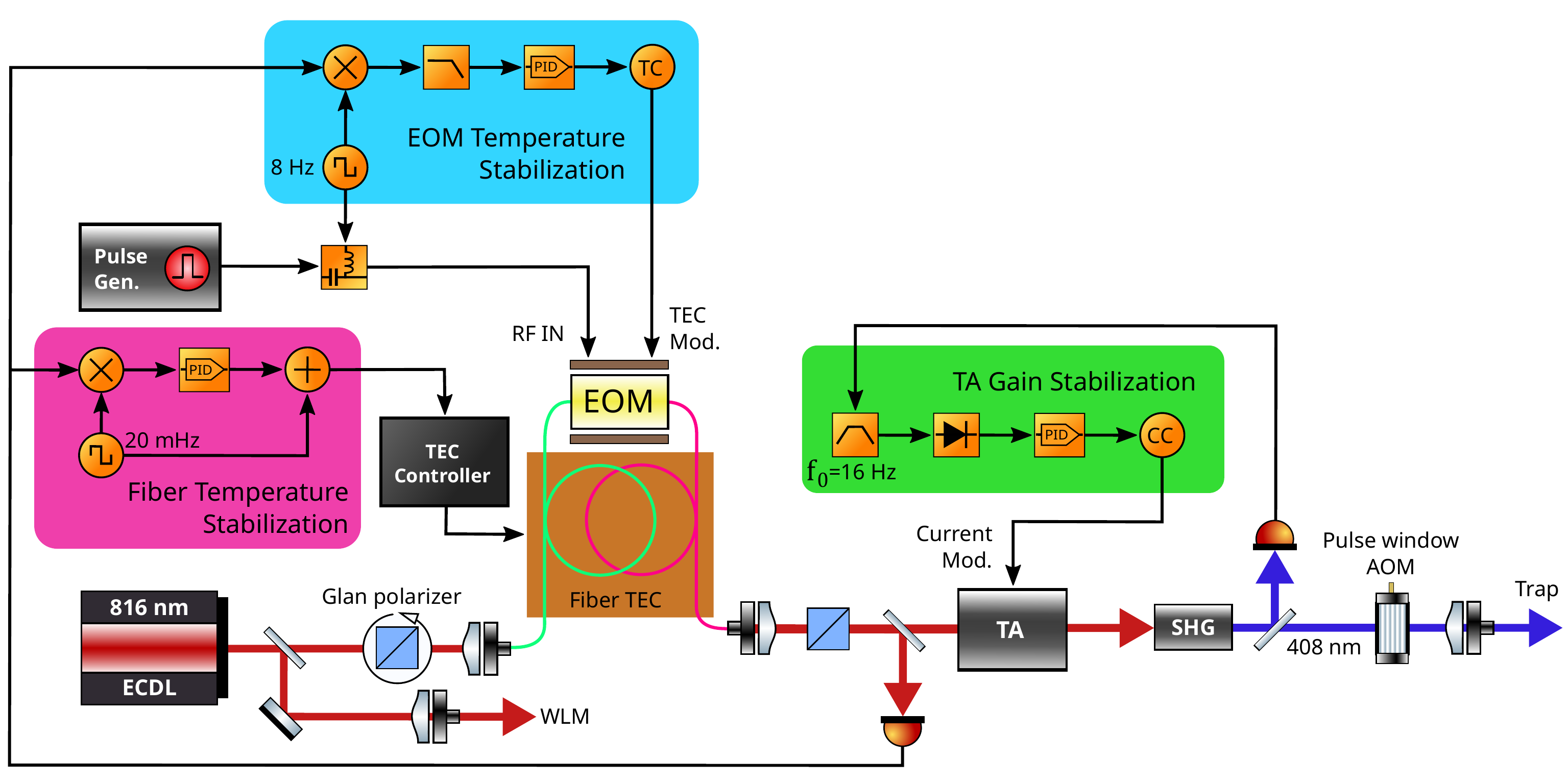}
    \caption{\label{fig:408nm_pulsed_system} Schematic of the custom 408\,nm pulsed laser system.}
\end{figure*}

To generate 408\,nm laser pulses, we use a CW 816\,nm custom ECDL built using a laser diode (Thorlabs L820P200), which is chopped by a fast electro-optic modulator (EOM) and then frequency doubled (Fig.~\ref{fig:408nm_pulsed_system}). The EOM (AdvR WPR-K0816) is a KTP-based waveguide device with an 18~pF input capacitance. It is used as an amplitude modulator, with a polarizer (Corning Polarcor) before the input fiber and a polarizing beamsplitter (Thorlabs PBS122) after the output fiber. The EOM is driven using a custom electrical pulse generation circuit based on a bias-switching step-recovery diode~\cite{Zou2017}. Further details and a circuit diagram can be found in Appendix~\ref{sec:AppendixA}. The output of the EOM is amplified by a TA (Coherent I14N-TA-0820-2000-DHP) and then frequency doubled using a fiber-coupled lithium niobate waveguide device (AdvR RSH-M0408). The doubled pulse is focused onto the ions via the optics used for optical pumping, with $\sigma^{-}$ polarization and a beam waist of approximately 20~{\textmu}m. We have implemented multiple stages of active stabilization to minimize optical power leakage through the EOM. These feedback systems are shown in the blue, green, and magenta blocks of Fig.~\ref{fig:408nm_pulsed_system} and are described in further detail in Appendix~\ref{sec:AppendixB}.

We measure the ratio between the peak optical pulse power and background light leakage when no electrical pulse is applied to the EOM. We record the average 408\,nm power at the doubler output with a power meter probe (Thorlabs S121C, specified with $\pm 3\,\%$ measurement uncertainty at 408\,nm) and vary the pulse repetition rate. Linear fitting produces a slope and intercept that correspond to a pulse energy of 19.1\,$\pm$\,0.2\,{\textmu}W/MHz = 19.1\,$\pm$\,0.2\,pJ and a CW background power of 1.92\,$\pm$\,0.04\,{\textmu}W, respectively. We then measure the pulse power versus time using a GaAs photodiode (EOT ET-4000) and an oscilloscope (Tektronix MSO 71254C) with 12.5\,GHz bandwidths (Fig.~\ref{fig:pulsed_system_output}). We collect a small fraction of the optical power onto the photodiode, and to find the peak power we vertically scale this trace such that the integral of the pulse trace equals the pulse energy found from the average power measurements. For a sample of seven 128-average traces, the standard deviation of the integrated pulse energy is 7\%. This gives a peak power of 52\,$\pm$\,4\,mW and extinction ratio $>$\,43\,dB.

The extinction ratio at the frequency doubler output is limited because the amplified spontaneous emission (ASE) from the TA is doubled. Due to the narrow phase-matching condition of the doubler, the spectrally broad ASE has low conversion efficiency. Despite this high attenuation, unwanted excitation of the ion would occur because of the low duty cycle. To mitigate this effect, we apply an 80\,ns gate using an acousto-optic modulator (Isomet M1250-T250L-0.45), increasing the final extinction ratio to 55\,dB.

\section{\label{sec:PhotonPurity}Single photon purity measurement}

\begin{figure}
    \centering
    \includegraphics[width=1.0\columnwidth]{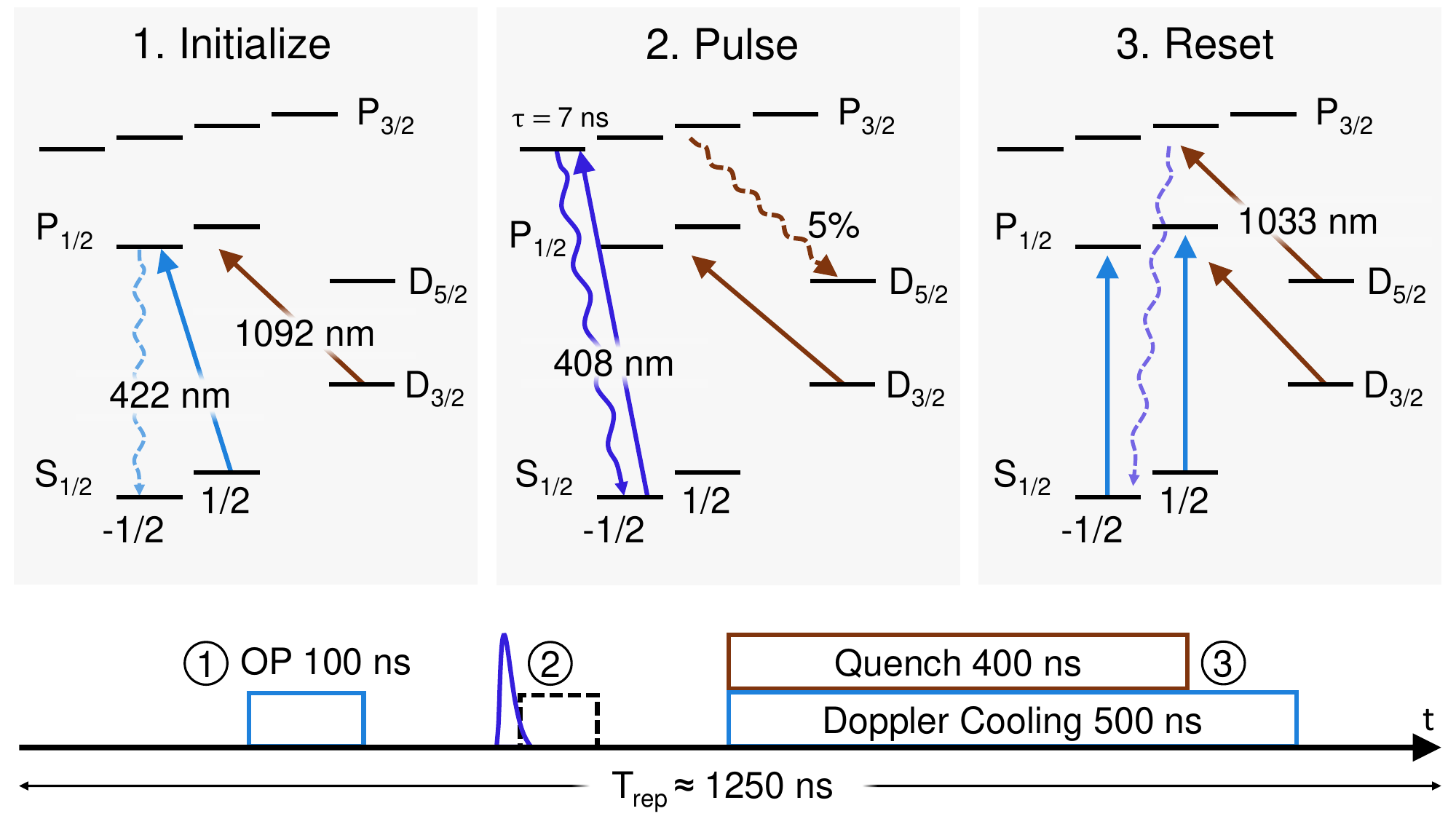}%
    \caption{\label{fig:energy_levels_and_g2_pulse_diagram} Energy levels of $^{88}$Sr$^+$ and pulse sequence used for $g^{(2)}$. The sequence includes state preparation, the excitation pulse, Doppler cooling, and clearing out the long-lived states of the ions, as described in the main text.}
\end{figure}

To characterize the single-photon purity, we measured the normalized second-order auto-correlation function $g^{(2)}(\tau)$ following the 408\,nm excitation pulse. A perfect single photon source exhibits $g^{(2)}(0)=0$, and any value of $g^{(2)}(0)<1$ implies suppression of multiphoton states compared to a classical light source~\cite{Loudon2000}. The experimental sequence for performing a $g^{(2)}(\tau)$ measurement starts by preparing the ion in the $\ket{\downarrow}$ state using 422\,nm $\sigma^-$-polarized light  (Fig.~\ref{fig:energy_levels_and_g2_pulse_diagram}). The ion is then excited using the $\sigma^-$-polarized 408\,nm excitation pulse. We collect the subsequent $P_{3/2}$ spontaneous emission into the HBT setup. A software gating window of 10\,ns is applied to each time-tagger channel approximately 4.5\,ns after the excitation pulse peak to exclude background counts arising from pulsed-laser light scattered from the ion trap and other surfaces. Then, since $P_{3/2}$ decay has approximately a 1:16 branching ratio to $D_{5/2}$~\cite{Zhang_iterative_2016}, a 1033\,nm quenching pulse removes any population trapped in $D_{5/2}$. Finally, we re-cool the ion with a 422\,nm Doppler cooling pulse. The 1092\,nm beam is applied continuously to clear out any population in $D_{3/2}$, since both $P_{3/2}$ and $P_{1/2}$ have branching ratios to this long-lived state~\cite{Zhang_iterative_2016}.

We use the Advanced Real-Time Infrastructure for Quantum physics (ARTIQ)~\cite{Bourdeauducq2021} framework and Sinara hardware family to orchestrate the experimental sequence, including controlling the AOMs via RF switches (Minicircuits ZASWA-2-50DRA+). Our ARTIQ build is modified to include the Oxford Ion Trap group's Entangler Core~\cite{EntanglerCore}, allowing the data acquisition sequence to operate at a $T_{rep}\approx1250$\,ns repetition period. 

\begin{figure}
    \centering
    \includegraphics[width=1.0\columnwidth]{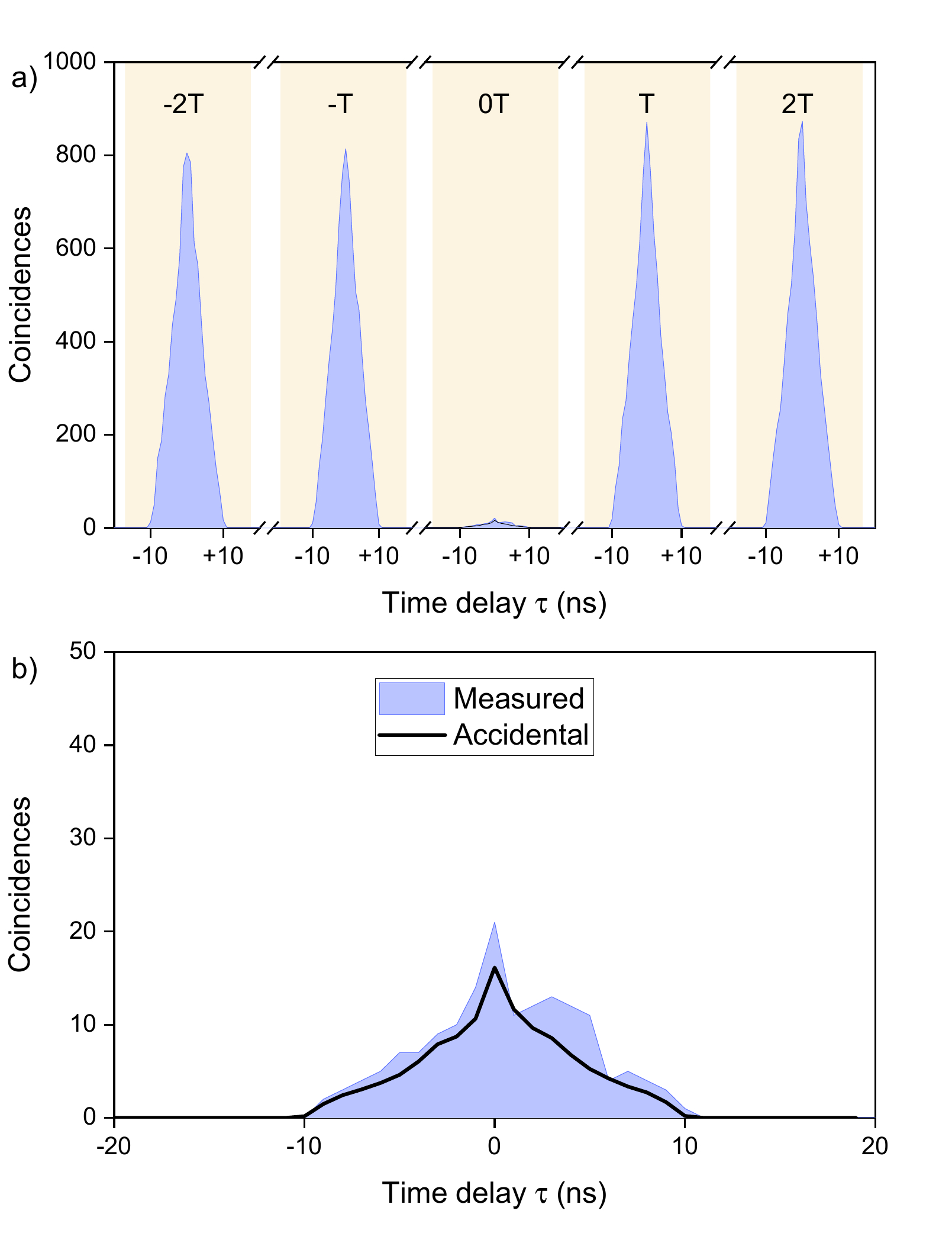}
    \caption{\label{fig:g2_one_ion} Correlation measurement for a single ion over 3 hours of applying the pulse sequence using 1\,ns bins. A 10\,ns gated window after the excitation pulse is applied to remove most of the scattered background light. (a) Coincidence distribution showing four side peaks with a temporal spacing of $T$ between each peak. (b) The coincidence distribution around $\tau=0$ is almost fully accounted for by the calculated contribution of the background (accidental coincidences), assuming perfect single photon emission from the ion.}
\end{figure}

The raw coincidence counts distribution from a single ion obtained over 3~hours is shown in Fig.~\ref{fig:g2_one_ion}(a) using 1\,ns bins. From the data, we can estimate the background-corrected $g^{(2)}(0)$ following Ref.~\citenum{hadfield_single_2007},
\begin{equation}
    \label{eq:g2_0}
    g^{(2)}(0)=\dfrac{C_0-C_B}{C_\tau-C_B},
\end{equation}
where $C_0$ and $C_\tau$ are the total number of coincidences integrated around $\tau=0$ and at peaks which form at multiples of the experiment repetition period (``side peaks''), respectively. The estimated mean background contribution is given by
\begin{equation}
    \label{eq:C_B}
    \langle C_B\rangle=T_{exp}T_{rep}(R_{T_1}R_{B_2}+R_{B_1}R_{T_2}-R_{B_1}R_{B_2}).
\end{equation}

\begin{table}
\caption{\label{tab:CBparams} Parameters used in Eq.~\ref{eq:C_B} for determining $g^{(2)}(0)$ for single-ion measurements. $R$ is the total singles count rate over the 10\,ns gated window (subscript $s$ and $b$ denote the ion fluorescence and background respectively, with the relevant PMT channel (CH) number labeled).}
\begin{ruledtabular}
\begin{tabular}{lll}
Parameter&Definition &Value\\
\hline
$T_{exp}$& Total experiment time& 3~hours\\
$T_{rep}$& Effective repetition period& $\approx$1250\,ns\\
$R_{T_1}$& Total count rate (CH 1)& 615.4\,$\pm$\,0.2\,s$^{-1}$\\
 $R_{B_1}$& Residual background count rate (CH 1)& 4.21\,$\pm$\,0.05\,s$^{-1}$\\
 $R_{T_2}$& Total count rate (CH 2)& 913.1\,$\pm$\,0.3\,s$^{-1}$\\
 $R_{B_2}$& Residual background count rate (CH 2)& 8.16\,$\pm$\,0.07\,s$^{-1}$\\
\end{tabular}
\end{ruledtabular}
\end{table}

The parameters in Eq.~\ref{eq:C_B} are defined in Table~\ref{tab:CBparams}. Over the total experiment time (3~hours), we record 158 coincidence counts at $\tau=0$ and an average of approximately 7700 coincidences in each of 32 side peaks. Additionally, the pulse sequence was applied for 30~minutes with no ion present to measure the residual background count rate on each channel. The coincidence counts distribution around $\tau=0$ is shown in Fig.~\ref{fig:g2_one_ion}(b). We integrate over the distributions around $\tau=0$ and across 32 side peaks to obtain the $g^{(2)}(0)$ value. From these results, we obtain $g^{(2)}(0)=(5.15\pm1.67) \times10^{-3}$, demonstrating our ability to produce single photons using the pulsed system. Note that, without background subtraction, $g^{(2)}(0)=(20.59\pm1.64)\times 10^{-3}$, demonstrating that the coincidences at $\tau=0$ in Fig.~\ref{fig:g2_one_ion} consist predominantly of background contributions. All uncertainties are statistical and reported as the standard error of the mean. The uncertainty in $C_\tau$ is estimated over 32 side peaks, and the uncertainty in $C_B$ is estimated using error propagation of the total and background singles counts measured.

We perform a similar experiment on a chain of two to six ions to show our ability to excite multiple ions at once, which may be useful for multiplexing schemes~\cite{PhysRevA.99.022337}. We apply the same experimental sequence and measure the total coincidences over 20 minutes around $\tau=0$ and across 16 side peaks. The residual background count rate was measured for 5~minutes. For $n$ independent emitters, the second-order correlation is given by~\cite{suarez_photon_2019}
\begin{equation}
    \label{eq:g2n_0}
    g^{(2)}_n(0)=1-\dfrac{1}{n}.
\end{equation}

\begin{figure}
    \centering
    \includegraphics[width=1.0\columnwidth]{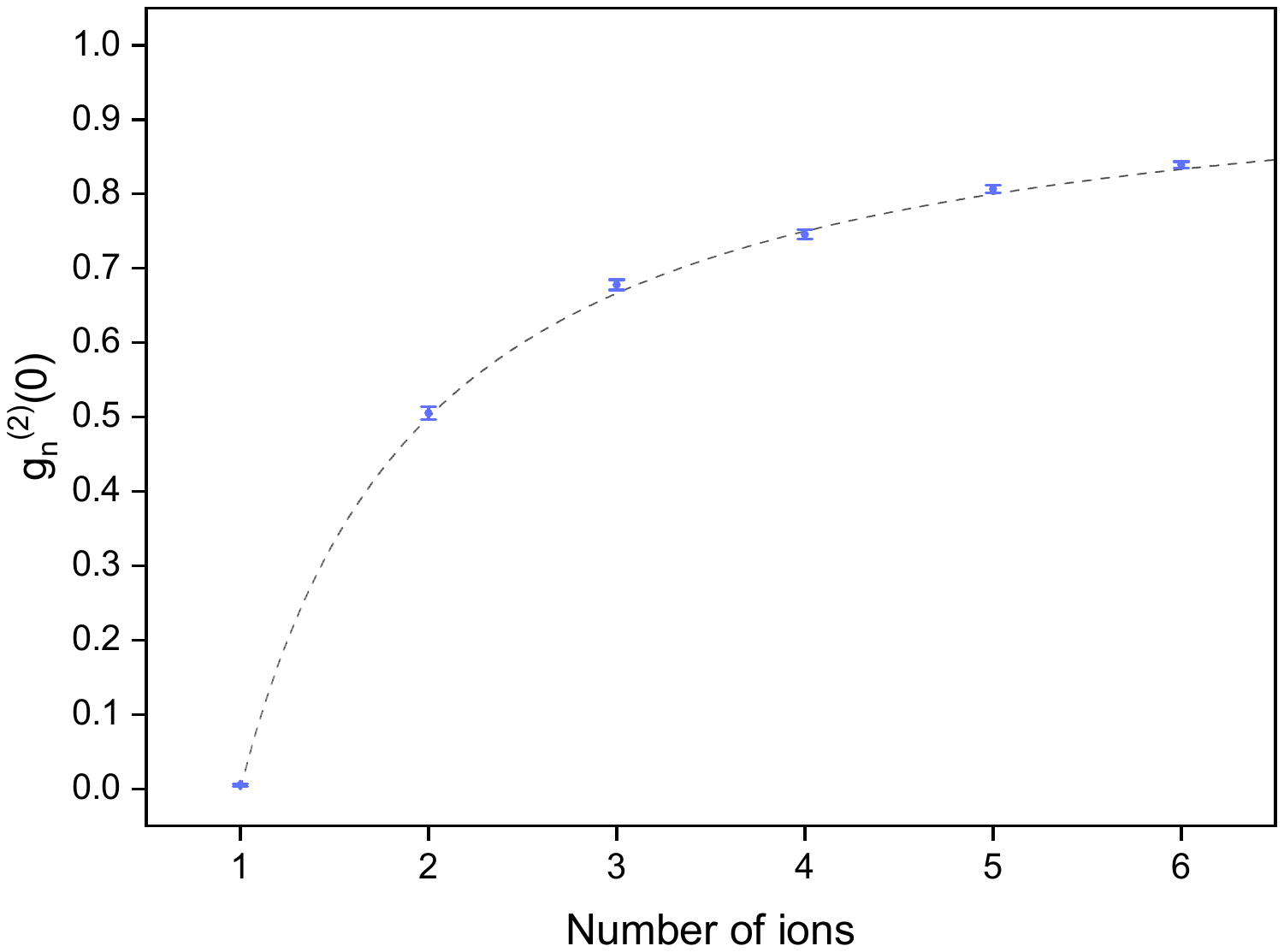}
    \caption{\label{fig:g2_many_ions} $g^{(2)}_n(0)$ as a function of ion number $n$. The $g^{(2)}$ value was calculated using Eq.~\ref{eq:g2_0} and is consistent with the prediction of $g_n^{(2)}(0)=1-1/n$ (dashed gray line) for $n$ independent emitters.}
\end{figure}

Fig.~\ref{fig:g2_many_ions} shows a plot of measured $g^{(2)}_n(0)$ versus ion number after background subtraction and is consistent with $n$ independent emission events per experimental cycle and uniform collection from the emitters. We note that most of the background contribution is due to scattered light from the excitation pulse. Assuming that the probability of $m$ emission events $P(m)$ from each ion is much smaller for multiple emission events compared to single events such that~\cite{Stevens_statistics_2013} $P(1)\gg\,P(2)\gg\,P(>2)$, $g^{(2)}(0)\approx2P(2)/P(1)^2$. An upper-bound on the mean infidelity due to multiple emission events can be approximated by $P(2)/P(1)\approx\frac{1}{2}P(1)g^{(2)}(0)=2.55\times10^{-3}$ for a single ion. This upper-bound may be achieved by better spatial or temporal filtering of the excitation pulse.

\section{\label{sec:Summary}Conclusion}

In this paper, we report the development of an instrument to trap $^{88}$Sr$^+$ ions and use them as a high-quality source of 408\,nm photons. We describe the vacuum and laser systems, including a custom 150\,ps pulsed laser system. We measured a $g^{(2)}(0)$ for a single ion that is competitive with state-of-the-art photon sources. For strings of multiple ions, we report a $g_n^{(2)}(0)$ consistent with independent emission by each ion. In the future, $^{88}$Sr$^+$ ions trapped in multiple zones may be ``remotely'' entangled by collecting and interfering 408\,nm light on the $S_{1/2} \leftrightarrow P_{3/2}$ transition. By implementing one- and two-qubit gates in each zone, this system would realize an architecture for exploring distributed quantum computing~\cite{Main_network_2025} and entanglement distillation~\cite{Kang_distillation_2023, Victora_distillation_2023}.

\clearpage

\begin{acknowledgments}
This work was supported by funding from the National Science Foundation through the Quantum Leap Challenge Institute for Hybrid Quantum Architectures and Networks (QLCI-HQAN) (award number 2016136). This material is also based upon work supported by the U.S. Department of Energy Office of Science National Quantum Information Science Research Centers. The Q-NEXT center supported the multi-ion $g_n^{(2)}(0)$ measurements. We acknowledge technical assistance from the MIT LL quantum computing group, including Felix Knollmann and Susanna Todaro.

\end{acknowledgments}

\section*{Author Declarations}

\subsection*{Conflict of Interest}

The authors have no conflicts to disclose.

\subsection*{Author Contributions}

\textbf{Jianlong Lin:} implemented the HBT setup and the experimental sequence for the $g^{(2)}$ measurement, measured and processed the data for single and multiple ions, and contributed to writing the manuscript. \textbf{Mari Cieszynski:} designed and built the vacuum system, integrated the surface trap and wiring, and contributed to writing the manuscript. \textbf{William Christopherson:} designed and built the laser, imaging, control, and ion loading systems and contributed to writing the manuscript. \textbf{Darman Khan:} contributed to taking and analyzing data for single and multiple ions. \textbf{Lintao Li:} designed and built the 674\,nm injection lock and pulsed laser systems. \textbf{Elizabeth Goldschmidt:} contributed to the design of the experiment and the writing of the manuscript. \textbf{Brian DeMarco:} contributed to the design of the experiment and the writing of the manuscript.

\section*{Data availability statement}
The data that support the findings of this study are available from the corresponding author upon reasonable request.

\appendix

\section{\label{sec:AppendixA}Pulse generation circuit}

\begin{figure}[ht!]
    \centering
    \includegraphics[width=1.0\columnwidth]{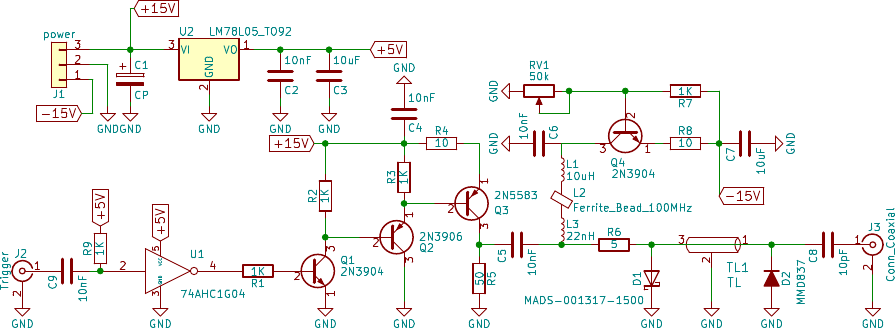} 
    \caption{\label{fig:pulse_circuit} Circuit used to generate voltage pulses applied to the 816\,nm EOM.}
\end{figure}

A schematic of the circuit that drives the 816\,nm EOM is shown in Fig.~\ref{fig:pulse_circuit}. The circuit uses the fast switching transition of a step-recovery diode (SRD) (Macom MMD837) to begin the pulse. The rising edge voltage is split, propagating to the output along a transmission line. After the transmission line, a short circuit inverts the voltage and it is reflected back along the transmission line. Finally, the inverted transition edge recombines with the initial edge to terminate the pulse. Before pulse generation, SRD D2 is forward biased by transistor Q4, which acts as a negative current source. Diode D1 is initially reverse biased. A pulse sent to the trigger input (J2) is amplified to 10\,V by transistors Q1, Q2, and Q3. The trigger pulse passes through C5 and reverse biases SRD D2, which switches with a transition time of 60\,ps and generates the rising edge of the pulse. The rising edge propagates to the output J3 and back along the transmission line TL1. D1 acts as a short to this positive voltage such that a negative-going reflection propagates back along TL1, thereby bringing the voltage at the output back to zero and terminating the output pulse. The transmission line length is set by the placement of the SRD on the printed circuit board and determines the pulse duration. For the pulse used in this work, TL1 is 1\,cm long.  

\section{\label{sec:AppendixB}408\,nm pulsed laser system feedback}

\subsection{\label{sec:EOMTempStab}816\,nm EOM temperature stabilization}

To achieve high-contrast pulses at 816\,nm, we separately stabilize the birefringence of the EOM and the input/output optical fibers via temperature control. Two independent feedback systems sample the light transmitted by the PBS that follows the EOM output fiber. The control systems are designed to minimize the transmitted optical power at zero applied voltage. They are able to operate continuously, since the low duty cycle during data collection results in an average 816\,nm power that is much smaller than the minimal static leakage. Furthermore, the experiment operates at a repetition rate that is incommensurate with the modulation frequency used for locking.

The EOM temperature and the optical fiber temperature are actively stabilized with feedback to two thermo-electric coolers (TECs): one mounted on the EOM body, and one which controls the temperature of the input/output fibers. The blue block of Fig.~\ref{fig:408nm_pulsed_system} shows how the control signal is generated and applied to the EOM TEC. A fraction of the output power is sent to a photodiode with a $1\,\mathrm{M\Omega}$ trans-impedance amplifier gain. Lock-in detection is used for stabilizing the temperature by introducing a small 8\,Hz modulation signal that is mixed with the photodiode signal and combined into the EOM AC modulation port through a bias-tee. 


Similar to the EOM temperature feedback system, we control the birefringence of the polarization-maintaining fibers by adjusting temperature (magenta block of Fig.~\ref{fig:408nm_pulsed_system}). The EOM input and output optical fibers are taped to a copper plate that is temperature controlled by a TEC. The frequency of fiber-temperature modulation is approximately 20\,mHz, which is sufficiently separated from the EOM modulation at 8\,Hz such that the two feedback stages are decoupled. Together, the stabilized system provides a 31\,dB extinction ratio, that is squared (on a linear scale) by the doubling stage's power threshold.


\subsection{\label{sec:TAGainStab}Tapered amplifier gain stabilization}

We have observed that the 408\,nm pulse shape and peak power are affected by the 816\,nm TA current. To optimize for peak power and minimum pulse length, we adjust the TA current based on a feedback system. This control system is designed to maximize the power of the 408\,nm light that results from frequency doubling the 816\,nm EOM leakage at zero applied voltage. The feedback system is able to operate continuously because of the low duty cycle of the 408\,nm pulse during data collection. 



For this feedback system, we make use of the modulation applied to the EOM voltage. As described in Appendix~\ref{sec:EOMTempStab}, an 8\,Hz modulation signal is applied to the EOM AC input. When the EOM temperature control systems are locked, this modulation signal is rectified in the EOM output power and doubled 408\,nm power such that the frequency is doubled to 16\,Hz. The green block in Fig.~\ref{fig:408nm_pulsed_system} outlines the TA gain feedback circuit that uses this 16\,Hz signal.

A fraction of the doubled 408\,nm light is picked off and measured by a photodiode with a $10\,\mathrm{M\Omega}$ transimpedence amplifier gain. The 16\,Hz signal has poor signal-to-noise ratio, so it is filtered using a band-pass. A PID controller uses this signal to control the TA gain by adjusting the TA current. Similar to the EOM temperature servo, this feedback loop is not affected by the pulse train as long as the experiment avoids a repetition rate commensurate with 16\,Hz.



\bibliography{g2paper}

\begin{thebibliography}{31}%
\makeatletter
\providecommand \@ifxundefined [1]{%
 \@ifx{#1\undefined}
}%
\providecommand \@ifnum [1]{%
 \ifnum #1\expandafter \@firstoftwo
 \else \expandafter \@secondoftwo
 \fi
}%
\providecommand \@ifx [1]{%
 \ifx #1\expandafter \@firstoftwo
 \else \expandafter \@secondoftwo
 \fi
}%
\providecommand \natexlab [1]{#1}%
\providecommand \enquote  [1]{``#1''}%
\providecommand \bibnamefont  [1]{#1}%
\providecommand \bibfnamefont [1]{#1}%
\providecommand \citenamefont [1]{#1}%
\providecommand \href@noop [0]{\@secondoftwo}%
\providecommand \href [0]{\begingroup \@sanitize@url \@href}%
\providecommand \@href[1]{\@@startlink{#1}\@@href}%
\providecommand \@@href[1]{\endgroup#1\@@endlink}%
\providecommand \@sanitize@url [0]{\catcode `\\12\catcode `\$12\catcode `\&12\catcode `\#12\catcode `\^12\catcode `\_12\catcode `\%12\relax}%
\providecommand \@@startlink[1]{}%
\providecommand \@@endlink[0]{}%
\providecommand \url  [0]{\begingroup\@sanitize@url \@url }%
\providecommand \@url [1]{\endgroup\@href {#1}{\urlprefix }}%
\providecommand \urlprefix  [0]{URL }%
\providecommand \Eprint [0]{\href }%
\providecommand \doibase [0]{http://dx.doi.org/}%
\providecommand \selectlanguage [0]{\@gobble}%
\providecommand \bibinfo  [0]{\@secondoftwo}%
\providecommand \bibfield  [0]{\@secondoftwo}%
\providecommand \translation [1]{[#1]}%
\providecommand \BibitemOpen [0]{}%
\providecommand \bibitemStop [0]{}%
\providecommand \bibitemNoStop [0]{.\EOS\space}%
\providecommand \EOS [0]{\spacefactor3000\relax}%
\providecommand \BibitemShut  [1]{\csname bibitem#1\endcsname}%
\let\auto@bib@innerbib\@empty
\bibitem [{\citenamefont {DiVincenzo}(2000)}]{DiVincenzo_computing_2000}%
  \BibitemOpen
  \bibfield  {author} {\bibinfo {author} {\bibfnamefont {D.~P.}\ \bibnamefont {DiVincenzo}},\ }\bibfield  {title} {\enquote {\bibinfo {title} {The physical implementation of quantum computation},}\ }\href {\doibase https://doi.org/10.1002/1521-3978(200009)48:9/11<771::AID-PROP771>3.0.CO;2-E} {\bibfield  {journal} {\bibinfo  {journal} {Fortschritte der Physik}\ }\textbf {\bibinfo {volume} {48}},\ \bibinfo {pages} {771--783} (\bibinfo {year} {2000})}\BibitemShut {NoStop}%
\bibitem [{\citenamefont {Kimble}(2008)}]{Kimble_internet_2008}%
  \BibitemOpen
  \bibfield  {author} {\bibinfo {author} {\bibfnamefont {H.~J.}\ \bibnamefont {Kimble}},\ }\bibfield  {title} {\enquote {\bibinfo {title} {The quantum internet},}\ }\href {\doibase 10.1038/nature07127} {\bibfield  {journal} {\bibinfo  {journal} {Nature}\ }\textbf {\bibinfo {volume} {453}},\ \bibinfo {pages} {1023--1030} (\bibinfo {year} {2008})}\BibitemShut {NoStop}%
\bibitem [{\citenamefont {Covey}, \citenamefont {Weinfurter},\ and\ \citenamefont {Bernien}(2023)}]{Covey_network_2023}%
  \BibitemOpen
  \bibfield  {author} {\bibinfo {author} {\bibfnamefont {J.~P.}\ \bibnamefont {Covey}}, \bibinfo {author} {\bibfnamefont {H.}~\bibnamefont {Weinfurter}}, \ and\ \bibinfo {author} {\bibfnamefont {H.}~\bibnamefont {Bernien}},\ }\bibfield  {title} {\enquote {\bibinfo {title} {Quantum networks with neutral atom processing nodes},}\ }\href {\doibase 10.1038/s41534-023-00759-9} {\bibfield  {journal} {\bibinfo  {journal} {npj Quantum Information}\ }\textbf {\bibinfo {volume} {9}},\ \bibinfo {pages} {90} (\bibinfo {year} {2023})}\BibitemShut {NoStop}%
\bibitem [{\citenamefont {Main}\ \emph {et~al.}(2025)\citenamefont {Main}, \citenamefont {Drmota}, \citenamefont {Nadlinger}, \citenamefont {Ainley}, \citenamefont {Agrawal}, \citenamefont {Nichol}, \citenamefont {Srinivas}, \citenamefont {Araneda},\ and\ \citenamefont {Lucas}}]{Main_network_2025}%
  \BibitemOpen
  \bibfield  {author} {\bibinfo {author} {\bibfnamefont {D.}~\bibnamefont {Main}}, \bibinfo {author} {\bibfnamefont {P.}~\bibnamefont {Drmota}}, \bibinfo {author} {\bibfnamefont {D.~P.}\ \bibnamefont {Nadlinger}}, \bibinfo {author} {\bibfnamefont {E.~M.}\ \bibnamefont {Ainley}}, \bibinfo {author} {\bibfnamefont {A.}~\bibnamefont {Agrawal}}, \bibinfo {author} {\bibfnamefont {B.~C.}\ \bibnamefont {Nichol}}, \bibinfo {author} {\bibfnamefont {R.}~\bibnamefont {Srinivas}}, \bibinfo {author} {\bibfnamefont {G.}~\bibnamefont {Araneda}}, \ and\ \bibinfo {author} {\bibfnamefont {D.~M.}\ \bibnamefont {Lucas}},\ }\bibfield  {title} {\enquote {\bibinfo {title} {Distributed quantum computing across an optical network link},}\ }\href {\doibase 10.1038/s41586-024-08404-x} {\bibfield  {journal} {\bibinfo  {journal} {Nature}\ }\textbf {\bibinfo {volume} {638}},\ \bibinfo {pages} {383--388} (\bibinfo {year} {2025})}\BibitemShut {NoStop}%
\bibitem [{\citenamefont {Flagg}\ \emph {et~al.}(2012)\citenamefont {Flagg}, \citenamefont {Polyakov}, \citenamefont {Thomay},\ and\ \citenamefont {Solomon}}]{flagg2012dynamics}%
  \BibitemOpen
  \bibfield  {author} {\bibinfo {author} {\bibfnamefont {E.~B.}\ \bibnamefont {Flagg}}, \bibinfo {author} {\bibfnamefont {S.~V.}\ \bibnamefont {Polyakov}}, \bibinfo {author} {\bibfnamefont {T.}~\bibnamefont {Thomay}}, \ and\ \bibinfo {author} {\bibfnamefont {G.~S.}\ \bibnamefont {Solomon}},\ }\bibfield  {title} {\enquote {\bibinfo {title} {Dynamics of nonclassical light from a single solid-state quantum emitter},}\ }\href@noop {} {\bibfield  {journal} {\bibinfo  {journal} {Physical review letters}\ }\textbf {\bibinfo {volume} {109}},\ \bibinfo {pages} {163601} (\bibinfo {year} {2012})}\BibitemShut {NoStop}%
\bibitem [{\citenamefont {Esmann}, \citenamefont {Wein},\ and\ \citenamefont {Antón-Solanas}(2024)}]{Esmann_solid_2024}%
  \BibitemOpen
  \bibfield  {author} {\bibinfo {author} {\bibfnamefont {M.}~\bibnamefont {Esmann}}, \bibinfo {author} {\bibfnamefont {S.~C.}\ \bibnamefont {Wein}}, \ and\ \bibinfo {author} {\bibfnamefont {C.}~\bibnamefont {Antón-Solanas}},\ }\bibfield  {title} {\enquote {\bibinfo {title} {Solid-state single-photon sources: Recent advances for novel quantum materials},}\ }\href {\doibase https://doi.org/10.1002/adfm.202315936} {\bibfield  {journal} {\bibinfo  {journal} {Advanced Functional Materials}\ }\textbf {\bibinfo {volume} {34}},\ \bibinfo {pages} {2315936} (\bibinfo {year} {2024})}\BibitemShut {NoStop}%
\bibitem [{\citenamefont {Liu}\ \emph {et~al.}(2021)\citenamefont {Liu}, \citenamefont {Cundiff}, \citenamefont {Almeida},\ and\ \citenamefont {Ulbricht}}]{Liu_nv_2021}%
  \BibitemOpen
  \bibfield  {author} {\bibinfo {author} {\bibfnamefont {A.}~\bibnamefont {Liu}}, \bibinfo {author} {\bibfnamefont {S.~T.}\ \bibnamefont {Cundiff}}, \bibinfo {author} {\bibfnamefont {D.~B.}\ \bibnamefont {Almeida}}, \ and\ \bibinfo {author} {\bibfnamefont {R.}~\bibnamefont {Ulbricht}},\ }\bibfield  {title} {\enquote {\bibinfo {title} {Spectral broadening and ultrafast dynamics of a nitrogen-vacancy center ensemble in diamond},}\ }\href {\doibase 10.1088/2633-4356/abf330} {\bibfield  {journal} {\bibinfo  {journal} {Materials for Quantum Technology}\ }\textbf {\bibinfo {volume} {1}},\ \bibinfo {pages} {025002} (\bibinfo {year} {2021})}\BibitemShut {NoStop}%
\bibitem [{\citenamefont {Andrini}\ \emph {et~al.}(2024)\citenamefont {Andrini}, \citenamefont {Amanti}, \citenamefont {Armani}, \citenamefont {Bellani}, \citenamefont {Bonaiuto}, \citenamefont {Cammarata}, \citenamefont {Campostrini}, \citenamefont {Dao}, \citenamefont {De~Matteis}, \citenamefont {Demontis}, \citenamefont {Di~Giuseppe}, \citenamefont {Ditalia~Tchernij}, \citenamefont {Donati}, \citenamefont {Fontana}, \citenamefont {Forneris}, \citenamefont {Francini}, \citenamefont {Frontini}, \citenamefont {Gunnella}, \citenamefont {Iadanza}, \citenamefont {Kaplan}, \citenamefont {Lacava}, \citenamefont {Liberali}, \citenamefont {Marzioni}, \citenamefont {Nieto~Hernández}, \citenamefont {Pedreschi}, \citenamefont {Piergentili}, \citenamefont {Prete}, \citenamefont {Prosposito}, \citenamefont {Rigato}, \citenamefont {Roncolato}, \citenamefont {Rossella}, \citenamefont {Salamon}, \citenamefont {Salvato}, \citenamefont {Sargeni}, \citenamefont {Shojaii}, \citenamefont {Spinella}, \citenamefont {Stabile},
  \citenamefont {Toncelli}, \citenamefont {Trucco},\ and\ \citenamefont {Vitali}}]{Andrini_photons_2024}%
  \BibitemOpen
  \bibfield  {author} {\bibinfo {author} {\bibfnamefont {G.}~\bibnamefont {Andrini}}, \bibinfo {author} {\bibfnamefont {F.}~\bibnamefont {Amanti}}, \bibinfo {author} {\bibfnamefont {F.}~\bibnamefont {Armani}}, \bibinfo {author} {\bibfnamefont {V.}~\bibnamefont {Bellani}}, \bibinfo {author} {\bibfnamefont {V.}~\bibnamefont {Bonaiuto}}, \bibinfo {author} {\bibfnamefont {S.}~\bibnamefont {Cammarata}}, \bibinfo {author} {\bibfnamefont {M.}~\bibnamefont {Campostrini}}, \bibinfo {author} {\bibfnamefont {T.~H.}\ \bibnamefont {Dao}}, \bibinfo {author} {\bibfnamefont {F.}~\bibnamefont {De~Matteis}}, \bibinfo {author} {\bibfnamefont {V.}~\bibnamefont {Demontis}}, \bibinfo {author} {\bibfnamefont {G.}~\bibnamefont {Di~Giuseppe}}, \bibinfo {author} {\bibfnamefont {S.}~\bibnamefont {Ditalia~Tchernij}}, \bibinfo {author} {\bibfnamefont {S.}~\bibnamefont {Donati}}, \bibinfo {author} {\bibfnamefont {A.}~\bibnamefont {Fontana}}, \bibinfo {author} {\bibfnamefont {J.}~\bibnamefont {Forneris}}, \bibinfo {author} {\bibfnamefont
  {R.}~\bibnamefont {Francini}}, \bibinfo {author} {\bibfnamefont {L.}~\bibnamefont {Frontini}}, \bibinfo {author} {\bibfnamefont {R.}~\bibnamefont {Gunnella}}, \bibinfo {author} {\bibfnamefont {S.}~\bibnamefont {Iadanza}}, \bibinfo {author} {\bibfnamefont {A.~E.}\ \bibnamefont {Kaplan}}, \bibinfo {author} {\bibfnamefont {C.}~\bibnamefont {Lacava}}, \bibinfo {author} {\bibfnamefont {V.}~\bibnamefont {Liberali}}, \bibinfo {author} {\bibfnamefont {F.}~\bibnamefont {Marzioni}}, \bibinfo {author} {\bibfnamefont {E.}~\bibnamefont {Nieto~Hernández}}, \bibinfo {author} {\bibfnamefont {E.}~\bibnamefont {Pedreschi}}, \bibinfo {author} {\bibfnamefont {P.}~\bibnamefont {Piergentili}}, \bibinfo {author} {\bibfnamefont {D.}~\bibnamefont {Prete}}, \bibinfo {author} {\bibfnamefont {P.}~\bibnamefont {Prosposito}}, \bibinfo {author} {\bibfnamefont {V.}~\bibnamefont {Rigato}}, \bibinfo {author} {\bibfnamefont {C.}~\bibnamefont {Roncolato}}, \bibinfo {author} {\bibfnamefont {F.}~\bibnamefont {Rossella}}, \bibinfo {author}
  {\bibfnamefont {A.}~\bibnamefont {Salamon}}, \bibinfo {author} {\bibfnamefont {M.}~\bibnamefont {Salvato}}, \bibinfo {author} {\bibfnamefont {F.}~\bibnamefont {Sargeni}}, \bibinfo {author} {\bibfnamefont {J.}~\bibnamefont {Shojaii}}, \bibinfo {author} {\bibfnamefont {F.}~\bibnamefont {Spinella}}, \bibinfo {author} {\bibfnamefont {A.}~\bibnamefont {Stabile}}, \bibinfo {author} {\bibfnamefont {A.}~\bibnamefont {Toncelli}}, \bibinfo {author} {\bibfnamefont {G.}~\bibnamefont {Trucco}}, \ and\ \bibinfo {author} {\bibfnamefont {V.}~\bibnamefont {Vitali}},\ }\bibfield  {title} {\enquote {\bibinfo {title} {Solid-state color centers for single-photon generation},}\ }\href {\doibase 10.3390/photonics11020188} {\bibfield  {journal} {\bibinfo  {journal} {Photonics}\ }\textbf {\bibinfo {volume} {11}} (\bibinfo {year} {2024}),\ 10.3390/photonics11020188}\BibitemShut {NoStop}%
\bibitem [{\citenamefont {Waltrich}\ \emph {et~al.}(2023)\citenamefont {Waltrich}, \citenamefont {Klotz}, \citenamefont {Agafonov},\ and\ \citenamefont {Kubanek}}]{Waltrich_interference_2023}%
  \BibitemOpen
  \bibfield  {author} {\bibinfo {author} {\bibfnamefont {R.}~\bibnamefont {Waltrich}}, \bibinfo {author} {\bibfnamefont {M.}~\bibnamefont {Klotz}}, \bibinfo {author} {\bibfnamefont {V.~N.}\ \bibnamefont {Agafonov}}, \ and\ \bibinfo {author} {\bibfnamefont {A.}~\bibnamefont {Kubanek}},\ }\bibfield  {title} {\enquote {\bibinfo {title} {Two-photon interference from silicon-vacancy centers in remote nanodiamonds},}\ }\href {\doibase doi:10.1515/nanoph-2023-0379} {\bibfield  {journal} {\bibinfo  {journal} {Nanophotonics}\ }\textbf {\bibinfo {volume} {12}},\ \bibinfo {pages} {3663--3669} (\bibinfo {year} {2023})}\BibitemShut {NoStop}%
\bibitem [{\citenamefont {Bruzewicz}\ \emph {et~al.}(2019)\citenamefont {Bruzewicz}, \citenamefont {Chiaverini}, \citenamefont {McConnell},\ and\ \citenamefont {Sage}}]{Bruzewicz_computing_2019}%
  \BibitemOpen
  \bibfield  {author} {\bibinfo {author} {\bibfnamefont {C.~D.}\ \bibnamefont {Bruzewicz}}, \bibinfo {author} {\bibfnamefont {J.}~\bibnamefont {Chiaverini}}, \bibinfo {author} {\bibfnamefont {R.}~\bibnamefont {McConnell}}, \ and\ \bibinfo {author} {\bibfnamefont {J.~M.}\ \bibnamefont {Sage}},\ }\bibfield  {title} {\enquote {\bibinfo {title} {Trapped-ion quantum computing: Progress and challenges},}\ }\href {\doibase 10.1063/1.5088164} {\bibfield  {journal} {\bibinfo  {journal} {Applied Physics Reviews}\ }\textbf {\bibinfo {volume} {6}},\ \bibinfo {pages} {021314} (\bibinfo {year} {2019})}\BibitemShut {NoStop}%
\bibitem [{\citenamefont {Sage}, \citenamefont {Kerman},\ and\ \citenamefont {Chiaverini}(2012)}]{sage_loading_2012}%
  \BibitemOpen
  \bibfield  {author} {\bibinfo {author} {\bibfnamefont {J.~M.}\ \bibnamefont {Sage}}, \bibinfo {author} {\bibfnamefont {A.~J.}\ \bibnamefont {Kerman}}, \ and\ \bibinfo {author} {\bibfnamefont {J.}~\bibnamefont {Chiaverini}},\ }\bibfield  {title} {\enquote {\bibinfo {title} {Loading of a surface-electrode ion trap from a remote, precooled source},}\ }\href {\doibase 10.1103/PhysRevA.86.013417} {\bibfield  {journal} {\bibinfo  {journal} {Physical Review A}\ }\textbf {\bibinfo {volume} {86}},\ \bibinfo {pages} {013417} (\bibinfo {year} {2012})}\BibitemShut {NoStop}%
\bibitem [{\citenamefont {Bruzewicz}\ \emph {et~al.}(2016)\citenamefont {Bruzewicz}, \citenamefont {McConnell}, \citenamefont {Chiaverini},\ and\ \citenamefont {Sage}}]{bruzewicz_scalable_2016}%
  \BibitemOpen
  \bibfield  {author} {\bibinfo {author} {\bibfnamefont {C.~D.}\ \bibnamefont {Bruzewicz}}, \bibinfo {author} {\bibfnamefont {R.}~\bibnamefont {McConnell}}, \bibinfo {author} {\bibfnamefont {J.}~\bibnamefont {Chiaverini}}, \ and\ \bibinfo {author} {\bibfnamefont {J.~M.}\ \bibnamefont {Sage}},\ }\bibfield  {title} {\enquote {\bibinfo {title} {Scalable loading of a two-dimensional trapped-ion array},}\ }\href {\doibase 10.1038/ncomms13005} {\bibfield  {journal} {\bibinfo  {journal} {Nature Communications}\ }\textbf {\bibinfo {volume} {7}},\ \bibinfo {pages} {13005} (\bibinfo {year} {2016})}\BibitemShut {NoStop}%
\bibitem [{\citenamefont {Huie}\ \emph {et~al.}(2023)\citenamefont {Huie}, \citenamefont {Li}, \citenamefont {Chen}, \citenamefont {Hu}, \citenamefont {Jia}, \citenamefont {Sun},\ and\ \citenamefont {Covey}}]{huie_repetitive_2023}%
  \BibitemOpen
  \bibfield  {author} {\bibinfo {author} {\bibfnamefont {W.}~\bibnamefont {Huie}}, \bibinfo {author} {\bibfnamefont {L.}~\bibnamefont {Li}}, \bibinfo {author} {\bibfnamefont {N.}~\bibnamefont {Chen}}, \bibinfo {author} {\bibfnamefont {X.}~\bibnamefont {Hu}}, \bibinfo {author} {\bibfnamefont {Z.}~\bibnamefont {Jia}}, \bibinfo {author} {\bibfnamefont {W.~K.~C.}\ \bibnamefont {Sun}}, \ and\ \bibinfo {author} {\bibfnamefont {J.~P.}\ \bibnamefont {Covey}},\ }\bibfield  {title} {\enquote {\bibinfo {title} {Repetitive readout and real-time control of nuclear spin qubits in \${\textasciicircum}\{171\}\${Yb} atoms},}\ }\href {\doibase 10.1103/PRXQuantum.4.030337} {\bibfield  {journal} {\bibinfo  {journal} {PRX Quantum}\ }\textbf {\bibinfo {volume} {4}},\ \bibinfo {pages} {030337} (\bibinfo {year} {2023})}\BibitemShut {NoStop}%
\bibitem [{\citenamefont {Chiaverini}\ and\ \citenamefont {Sage}(2014)}]{PhysRevA.89.012318}%
  \BibitemOpen
  \bibfield  {author} {\bibinfo {author} {\bibfnamefont {J.}~\bibnamefont {Chiaverini}}\ and\ \bibinfo {author} {\bibfnamefont {J.~M.}\ \bibnamefont {Sage}},\ }\bibfield  {title} {\enquote {\bibinfo {title} {Insensitivity of the rate of ion motional heating to trap-electrode material over a large temperature range},}\ }\href {\doibase 10.1103/PhysRevA.89.012318} {\bibfield  {journal} {\bibinfo  {journal} {Phys. Rev. A}\ }\textbf {\bibinfo {volume} {89}},\ \bibinfo {pages} {012318} (\bibinfo {year} {2014})}\BibitemShut {NoStop}%
\bibitem [{\citenamefont {Shi}\ \emph {et~al.}(2024)\citenamefont {Shi}, \citenamefont {Sinanan-Singh}, \citenamefont {DeBry}, \citenamefont {Todaro}, \citenamefont {Chuang},\ and\ \citenamefont {Chiaverini}}]{shi_long-lived_2024}%
  \BibitemOpen
  \bibfield  {author} {\bibinfo {author} {\bibfnamefont {X.}~\bibnamefont {Shi}}, \bibinfo {author} {\bibfnamefont {J.}~\bibnamefont {Sinanan-Singh}}, \bibinfo {author} {\bibfnamefont {K.}~\bibnamefont {DeBry}}, \bibinfo {author} {\bibfnamefont {S.~L.}\ \bibnamefont {Todaro}}, \bibinfo {author} {\bibfnamefont {I.~L.}\ \bibnamefont {Chuang}}, \ and\ \bibinfo {author} {\bibfnamefont {J.}~\bibnamefont {Chiaverini}},\ }\bibfield  {title} {\enquote {\bibinfo {title} {Long-lived metastable-qubit memory},}\ }\href@noop {} {\  (\bibinfo {year} {2024})},\ \bibinfo {note} {\_eprint: 2408.00975}\BibitemShut {NoStop}%
\bibitem [{\citenamefont {Keselman}\ \emph {et~al.}(2011)\citenamefont {Keselman}, \citenamefont {Glickman}, \citenamefont {Akerman}, \citenamefont {Kotler},\ and\ \citenamefont {Ozeri}}]{Keselman_state_detection_2011}%
  \BibitemOpen
  \bibfield  {author} {\bibinfo {author} {\bibfnamefont {A.}~\bibnamefont {Keselman}}, \bibinfo {author} {\bibfnamefont {Y.}~\bibnamefont {Glickman}}, \bibinfo {author} {\bibfnamefont {N.}~\bibnamefont {Akerman}}, \bibinfo {author} {\bibfnamefont {S.}~\bibnamefont {Kotler}}, \ and\ \bibinfo {author} {\bibfnamefont {R.}~\bibnamefont {Ozeri}},\ }\bibfield  {title} {\enquote {\bibinfo {title} {High-fidelity state detection and tomography of a single-ion zeeman qubit},}\ }\href {\doibase 10.1088/1367-2630/13/7/073027} {\bibfield  {journal} {\bibinfo  {journal} {New Journal of Physics}\ }\textbf {\bibinfo {volume} {13}},\ \bibinfo {pages} {073027} (\bibinfo {year} {2011})}\BibitemShut {NoStop}%
\bibitem [{\citenamefont {Shiner}(2006)}]{Shiner2006}%
  \BibitemOpen
  \bibfield  {author} {\bibinfo {author} {\bibfnamefont {A.}~\bibnamefont {Shiner}},\ }\emph {\bibinfo {title} {Development of a Frequency Stabilized 422-nm Diode-laser System and Its Application to a Strontium-88 Single Ion Optical Frequency Standard}},\ \href@noop {} {\bibinfo {type} {mathesis}},\ \bibinfo  {school} {York University} (\bibinfo {year} {2006})\BibitemShut {NoStop}%
\bibitem [{\citenamefont {Knollman}\ \emph {et~al.}(2024)\citenamefont {Knollman}, \citenamefont {Clements}, \citenamefont {DeBry}, \citenamefont {Todaro}, \citenamefont {Chuang},\ and\ \citenamefont {Chiaverini}}]{beam_delivery_MIT_2024}%
  \BibitemOpen
  \bibfield  {author} {\bibinfo {author} {\bibfnamefont {F.}~\bibnamefont {Knollman}}, \bibinfo {author} {\bibfnamefont {E.}~\bibnamefont {Clements}}, \bibinfo {author} {\bibfnamefont {K.}~\bibnamefont {DeBry}}, \bibinfo {author} {\bibfnamefont {S.~L.}\ \bibnamefont {Todaro}}, \bibinfo {author} {\bibfnamefont {I.~L.}\ \bibnamefont {Chuang}}, \ and\ \bibinfo {author} {\bibfnamefont {J.}~\bibnamefont {Chiaverini}},\ }\bibfield  {title} {\enquote {\bibinfo {title} {Private communication},}\ }\href@noop {} {\  (\bibinfo {year} {2024})},\ \bibinfo {note} {\_eprint: 2408.00975}\BibitemShut {NoStop}%
\bibitem [{\citenamefont {Sansonetti}(2012)}]{Sansonetti2012}%
  \BibitemOpen
  \bibfield  {author} {\bibinfo {author} {\bibfnamefont {J.~E.}\ \bibnamefont {Sansonetti}},\ }\bibfield  {title} {\enquote {\bibinfo {title} {Wavelengths, transition probabilities, and energy levels for the spectra of strontium ions (sr ii through sr xxxviii)},}\ }\href {\doibase 10.1063/1.3659413} {\bibfield  {journal} {\bibinfo  {journal} {Journal of Physical and Chemical Reference Data}\ }\textbf {\bibinfo {volume} {41}},\ \bibinfo {pages} {013102--013102--119} (\bibinfo {year} {2012})}\BibitemShut {NoStop}%
\bibitem [{\citenamefont {Brownnutt}(2007)}]{Brownnutt2007}%
  \BibitemOpen
  \bibfield  {author} {\bibinfo {author} {\bibfnamefont {M.}~\bibnamefont {Brownnutt}},\ }\emph {\bibinfo {title} {88Sr+ ion trapping techniques and technologies for quantum information processing}},\ \href@noop {} {Ph.D. thesis},\ \bibinfo  {school} {Imperial College London} (\bibinfo {year} {2007})\BibitemShut {NoStop}%
\bibitem [{\citenamefont {Zou}, \citenamefont {Gupta},\ and\ \citenamefont {Caloz}(2017)}]{Zou2017}%
  \BibitemOpen
  \bibfield  {author} {\bibinfo {author} {\bibfnamefont {L.}~\bibnamefont {Zou}}, \bibinfo {author} {\bibfnamefont {S.}~\bibnamefont {Gupta}}, \ and\ \bibinfo {author} {\bibfnamefont {C.}~\bibnamefont {Caloz}},\ }\bibfield  {title} {\enquote {\bibinfo {title} {A simple picosecond pulse generator based on a pair of step recovery diodes},}\ }\href {\doibase 10.1109/lmwc.2017.2690880} {\bibfield  {journal} {\bibinfo  {journal} {IEEE Microwave and Wireless Components Letters}\ }\textbf {\bibinfo {volume} {27}},\ \bibinfo {pages} {467--469} (\bibinfo {year} {2017})}\BibitemShut {NoStop}%
\bibitem [{\citenamefont {Loudon}(2000)}]{Loudon2000}%
  \BibitemOpen
  \bibfield  {author} {\bibinfo {author} {\bibfnamefont {R.}~\bibnamefont {Loudon}},\ }\href {\doibase 10.1093/oso/9780198501770.001.0001} {\emph {\bibinfo {title} {{The Quantum Theory of Light}}}}\ (\bibinfo  {publisher} {Oxford University Press},\ \bibinfo {year} {2000})\BibitemShut {NoStop}%
\bibitem [{\citenamefont {Zhang}\ \emph {et~al.}(2016)\citenamefont {Zhang}, \citenamefont {Gutierrez}, \citenamefont {Low}, \citenamefont {Rines}, \citenamefont {Stuart}, \citenamefont {Wu},\ and\ \citenamefont {Chuang}}]{Zhang_iterative_2016}%
  \BibitemOpen
  \bibfield  {author} {\bibinfo {author} {\bibfnamefont {H.}~\bibnamefont {Zhang}}, \bibinfo {author} {\bibfnamefont {M.}~\bibnamefont {Gutierrez}}, \bibinfo {author} {\bibfnamefont {G.~H.}\ \bibnamefont {Low}}, \bibinfo {author} {\bibfnamefont {R.}~\bibnamefont {Rines}}, \bibinfo {author} {\bibfnamefont {J.}~\bibnamefont {Stuart}}, \bibinfo {author} {\bibfnamefont {T.}~\bibnamefont {Wu}}, \ and\ \bibinfo {author} {\bibfnamefont {I.}~\bibnamefont {Chuang}},\ }\bibfield  {title} {\enquote {\bibinfo {title} {Iterative precision measurement of branching ratios applied to 5p states in 88sr+},}\ }\href {\doibase 10.1088/1367-2630/aa511d} {\bibfield  {journal} {\bibinfo  {journal} {New Journal of Physics}\ }\textbf {\bibinfo {volume} {18}},\ \bibinfo {pages} {123021} (\bibinfo {year} {2016})}\BibitemShut {NoStop}%
\bibitem [{\citenamefont {Bourdeauducq}\ \emph {et~al.}(2021)\citenamefont {Bourdeauducq}, \citenamefont {{Whitequark}}, \citenamefont {Jördens}, \citenamefont {Nadlinger}, \citenamefont {Sionneau},\ and\ \citenamefont {Kermarrec}}]{Bourdeauducq2021}%
  \BibitemOpen
  \bibfield  {author} {\bibinfo {author} {\bibfnamefont {S.}~\bibnamefont {Bourdeauducq}}, \bibinfo {author} {\bibnamefont {{Whitequark}}}, \bibinfo {author} {\bibfnamefont {R.}~\bibnamefont {Jördens}}, \bibinfo {author} {\bibfnamefont {D.}~\bibnamefont {Nadlinger}}, \bibinfo {author} {\bibfnamefont {Y.}~\bibnamefont {Sionneau}}, \ and\ \bibinfo {author} {\bibfnamefont {F.}~\bibnamefont {Kermarrec}},\ }\href {\doibase 10.5281/ZENODO.6619071} {\enquote {\bibinfo {title} {Artiq},}\ } (\bibinfo {year} {2021})\BibitemShut {NoStop}%
\bibitem [{\citenamefont {Risinger}(2021)}]{EntanglerCore}%
  \BibitemOpen
  \bibfield  {author} {\bibinfo {author} {\bibfnamefont {D.}~\bibnamefont {Risinger}},\ }\href@noop {} {\enquote {\bibinfo {title} {Entangler core},}\ }\bibinfo {howpublished} {\url{https://github.com/OxfordIonTrapGroup/entangler-core}} (\bibinfo {year} {2021})\BibitemShut {NoStop}%
\bibitem [{\citenamefont {Hadfield}\ \emph {et~al.}(2007)\citenamefont {Hadfield}, \citenamefont {Stevens}, \citenamefont {Mirin},\ and\ \citenamefont {Nam}}]{hadfield_single_2007}%
  \BibitemOpen
  \bibfield  {author} {\bibinfo {author} {\bibfnamefont {R.~H.}\ \bibnamefont {Hadfield}}, \bibinfo {author} {\bibfnamefont {M.~J.}\ \bibnamefont {Stevens}}, \bibinfo {author} {\bibfnamefont {R.~P.}\ \bibnamefont {Mirin}}, \ and\ \bibinfo {author} {\bibfnamefont {S.~W.}\ \bibnamefont {Nam}},\ }\bibfield  {title} {\enquote {\bibinfo {title} {{Single-photon source characterization with twin infrared-sensitive superconducting single-photon detectors}},}\ }\href {\doibase 10.1063/1.2717582} {\bibfield  {journal} {\bibinfo  {journal} {Journal of Applied Physics}\ }\textbf {\bibinfo {volume} {101}},\ \bibinfo {pages} {103104} (\bibinfo {year} {2007})}\BibitemShut {NoStop}%
\bibitem [{\citenamefont {Lo~Piparo}, \citenamefont {Munro},\ and\ \citenamefont {Nemoto}(2019)}]{PhysRevA.99.022337}%
  \BibitemOpen
  \bibfield  {author} {\bibinfo {author} {\bibfnamefont {N.}~\bibnamefont {Lo~Piparo}}, \bibinfo {author} {\bibfnamefont {W.~J.}\ \bibnamefont {Munro}}, \ and\ \bibinfo {author} {\bibfnamefont {K.}~\bibnamefont {Nemoto}},\ }\bibfield  {title} {\enquote {\bibinfo {title} {Quantum multiplexing},}\ }\href {\doibase 10.1103/PhysRevA.99.022337} {\bibfield  {journal} {\bibinfo  {journal} {Phys. Rev. A}\ }\textbf {\bibinfo {volume} {99}},\ \bibinfo {pages} {022337} (\bibinfo {year} {2019})}\BibitemShut {NoStop}%
\bibitem [{\citenamefont {Suarez}\ \emph {et~al.}(2019)\citenamefont {Suarez}, \citenamefont {Auwärter}, \citenamefont {Arruda}, \citenamefont {Bachelard}, \citenamefont {Courteille}, \citenamefont {Zimmermann},\ and\ \citenamefont {Slama}}]{suarez_photon_2019}%
  \BibitemOpen
  \bibfield  {author} {\bibinfo {author} {\bibfnamefont {E.}~\bibnamefont {Suarez}}, \bibinfo {author} {\bibfnamefont {D.}~\bibnamefont {Auwärter}}, \bibinfo {author} {\bibfnamefont {T.~J.}\ \bibnamefont {Arruda}}, \bibinfo {author} {\bibfnamefont {R.}~\bibnamefont {Bachelard}}, \bibinfo {author} {\bibfnamefont {P.~W.}\ \bibnamefont {Courteille}}, \bibinfo {author} {\bibfnamefont {C.}~\bibnamefont {Zimmermann}}, \ and\ \bibinfo {author} {\bibfnamefont {S.}~\bibnamefont {Slama}},\ }\bibfield  {title} {\enquote {\bibinfo {title} {Photon-antibunching in the fluorescence of statistical ensembles of emitters at an optical nanofiber-tip},}\ }\href {\doibase 10.1088/1367-2630/ab0a99} {\bibfield  {journal} {\bibinfo  {journal} {New Journal of Physics}\ }\textbf {\bibinfo {volume} {21}},\ \bibinfo {pages} {035009} (\bibinfo {year} {2019})}\BibitemShut {NoStop}%
\bibitem [{\citenamefont {Stevens}(2013)}]{Stevens_statistics_2013}%
  \BibitemOpen
  \bibfield  {author} {\bibinfo {author} {\bibfnamefont {M.~J.}\ \bibnamefont {Stevens}},\ }\bibfield  {title} {\enquote {\bibinfo {title} {Chapter 2 - photon statistics, measurements, and measurements tools},}\ }in\ \href {\doibase https://doi.org/10.1016/B978-0-12-387695-9.00002-0} {\emph {\bibinfo {booktitle} {Single-Photon Generation and Detection}}},\ \bibinfo {series} {Experimental Methods in the Physical Sciences}, Vol.~\bibinfo {volume} {45},\ \bibinfo {editor} {edited by\ \bibinfo {editor} {\bibfnamefont {A.}~\bibnamefont {Migdall}}, \bibinfo {editor} {\bibfnamefont {S.~V.}\ \bibnamefont {Polyakov}}, \bibinfo {editor} {\bibfnamefont {J.}~\bibnamefont {Fan}}, \ and\ \bibinfo {editor} {\bibfnamefont {J.~C.}\ \bibnamefont {Bienfang}}}\ (\bibinfo  {publisher} {Academic Press},\ \bibinfo {year} {2013})\ pp.\ \bibinfo {pages} {25--68}\BibitemShut {NoStop}%
\bibitem [{\citenamefont {Kang}\ \emph {et~al.}(2023)\citenamefont {Kang}, \citenamefont {Guha}, \citenamefont {Rengaswamy},\ and\ \citenamefont {Seshadreesan}}]{Kang_distillation_2023}%
  \BibitemOpen
  \bibfield  {author} {\bibinfo {author} {\bibfnamefont {A.}~\bibnamefont {Kang}}, \bibinfo {author} {\bibfnamefont {S.}~\bibnamefont {Guha}}, \bibinfo {author} {\bibfnamefont {N.}~\bibnamefont {Rengaswamy}}, \ and\ \bibinfo {author} {\bibfnamefont {K.~P.}\ \bibnamefont {Seshadreesan}},\ }\bibfield  {title} {\enquote {\bibinfo {title} {Trapped ion quantum repeaters with entanglement distillation based on quantum ldpc codes},}\ }in\ \href {\doibase 10.1109/QCE57702.2023.00131} {\emph {\bibinfo {booktitle} {2023 IEEE International Conference on Quantum Computing and Engineering (QCE)}}},\ Vol.~\bibinfo {volume} {01}\ (\bibinfo {year} {2023})\ pp.\ \bibinfo {pages} {1165--1171}\BibitemShut {NoStop}%
\bibitem [{\citenamefont {Victora}\ \emph {et~al.}(2023)\citenamefont {Victora}, \citenamefont {Tserkis}, \citenamefont {Krastanov}, \citenamefont {de~la Cerda}, \citenamefont {Willis},\ and\ \citenamefont {Narang}}]{Victora_distillation_2023}%
  \BibitemOpen
  \bibfield  {author} {\bibinfo {author} {\bibfnamefont {M.}~\bibnamefont {Victora}}, \bibinfo {author} {\bibfnamefont {S.}~\bibnamefont {Tserkis}}, \bibinfo {author} {\bibfnamefont {S.}~\bibnamefont {Krastanov}}, \bibinfo {author} {\bibfnamefont {A.~S.}\ \bibnamefont {de~la Cerda}}, \bibinfo {author} {\bibfnamefont {S.}~\bibnamefont {Willis}}, \ and\ \bibinfo {author} {\bibfnamefont {P.}~\bibnamefont {Narang}},\ }\bibfield  {title} {\enquote {\bibinfo {title} {Entanglement purification on quantum networks},}\ }\href {\doibase 10.1103/PhysRevResearch.5.033171} {\bibfield  {journal} {\bibinfo  {journal} {Phys. Rev. Res.}\ }\textbf {\bibinfo {volume} {5}},\ \bibinfo {pages} {033171} (\bibinfo {year} {2023})}\BibitemShut {NoStop}%
\end{thebibliography}%

\end{document}